\documentclass[article]{jss}

\usepackage[english]{babel}
\usepackage{blindtext}

\usepackage{verbatim}
\usepackage{amsmath}
\usepackage{amssymb}
\usepackage{algorithm2e}

\definecolor{tomato}{RGB}{255,99,71}
\definecolor{jssblue}{RGB}{32, 114, 247}

\usepackage{bm}
\usepackage{tikz}
\usetikzlibrary{arrows,shapes,trees} 
\usepackage{float}

\newcommand\scalemath[2]{\scalebox{#1}{\mbox{\ensuremath{\displaystyle #2}}}}

\newcommand\independent{\protect\mathpalette{\protect\independenT}{\perp}} 
\def\independenT#1#2{\mathrel{\rlap{$#1#2$}\mkern2mu{#1#2}}} 

\author{Jonas M. B. Haslbeck\\ University of Amsterdam \And 
        Lourens J. Waldorp\\ University of Amsterdam}
\title{\pkg{mgm}: Estimating Time-Varying Mixed Graphical Models in High-Dimensional Data}

\Plainauthor{Jonas Haslbeck, Lourens Waldorp} 
\Plaintitle{\pkg{mgm}: Estimating Time-Varying Mixed Graphical Models in High-Dimensional Data} 
\Shorttitle{\pkg{mgm}: Estimating Time-Varying Mixed Graphical Models in High-Dimensional Data} 

\Abstract{
We present the \proglang{R}-package \pkg{mgm} for the estimation of $k$-order Mixed Graphical Models (MGMs) and mixed Vector Autoregressive (mVAR) models in high-dimensional data. These are a useful extensions of graphical models for only one variable type, since data sets consisting of mixed types of variables (continuous, count, categorical) are ubiquitous. In addition, we allow to relax the stationarity assumption of both models by introducing time-varying versions MGMs and mVAR models based on a kernel weighting approach. Time-varying models offer a rich description of temporally evolving systems and allow to identify external influences on the model structure such as the impact of interventions. We provide the background of all implemented methods and provide fully reproducible examples that illustrate how to use the package.}


\Plainkeywords{structure estimation, mixed graphical models, Markov Random Fields, dynamic graphical models, time-varying graphical models, mixed vector autoregressive models} 

\Address{
Jonas Haslbeck\\
Psychological Methods\\
Nieuwe Achtergracht 129-B\\
Postbus 15906\\
1018 WT, Amsterdam\\
The Netherlands\\
E-Mail: \email{jonashaslbeck@gmail.com}\\
Website: \url{www.jonashaslbeck.com}
  }

\begin{document}

\section[Introduction]{Introduction}

We present \pkg{mgm}, an \proglang{R}-package for the estimation of (time-varying) $k$-order Mixed Graphical Models (MGMs) and (time-varying) mixed Vector Autoregressive (mVAR) models with a specified set of lags. The package is available from the Comprehensive R Archive Network (CRAN) at \url{http://CRAN.r-project.org/}. In this paper we introduce these models, discuss algorithms to estimate them, and present a number of fully reproducible code examples that show how to use the implementations provided by \pkg{mgm}.

Graphical models have become a popular way to abstract complex systems and gain insights into relational patterns among observed variables in a large variety of disciplines such as statistical mechanics \citep{Albert_statistical_2002}, biology \citep{friedmanusing2000}, genetics \citep{Ghazalpour_integrating_2006}, neuroscience \citep{huang_learning_2010} and psychology \citep{Borsboom_Network_2013}. In many of these applications the dataset of interest consists of \textit{mixed variables} such as binary, categorical, ordinal, counts, continuous and/or skewed continuous amongst others. One example is internet-scale marketing data, where it is of interest to relate variables such as clicked links (categorical), time spent on websites (possibly exponential), browsing history (categorical), social media postings (count), friends in social networks (count), and many others. In a medical context, one could be interested in interactions between person characteristics such as gender (categorical) or age (continuous), frequencies of behaviors (count), taking place of events (categorical) and the dose of a drug (continuous). 

If measurements are taken repeatedly from a system, one can be either interested in relations between variables at the \emph{same time point} or in relations between variables \emph{across time points}. The former relations are modeled by MGMs, the latter relations are modeled by Vector Autoregressive (VAR) models, which relate variables over a specified set of time lags. For both types of models it may be appropriate in some situations to relax the assumption of stationarity, such that its parameters are allowed to vary over the measured time period. These time-varying models provide additional information for understanding and predicting organizational processes, the diffusion of information, detecting vulnerabilities and the potential impact of interventions. An example is the developmental cycle of a biological organism, in which different genes interact at different stages of development. In a medical context, the aim could be to study the impact of an intervention on the dependencies between a large number of physiological and psychological variables that capture the health of a patient. Yet another example can be found in the field of psychiatry, where one might be interested in the interaction of negative life events, social contacts and symptoms of psychological disorders such as major depression.

\subsection{Implementation and functionality}

The \pkg{mgm} package is written in \proglang{R} and uses the \pkg{glmnet} package \citep{friedman2010regularization} to fit penalized Generalized Linear Models (GLMs) to perform neighborhood selection \citep{meinshausen_high-dimensional_2006}. The  \pkg{glmnet} package is written in \proglang{Fortran} and is optimized for computational efficiency. In addition, \pkg{mgm} depends on the packages \pkg{matrixcalc}, \pkg{stringr}, \pkg{Hmisc}, \pkg{gtools} and \pkg{qgraph}.

The main functionality of the \pkg{mgm} package is to estimate Mixed Graphical Models (MGMs) and mixed Autoregressive (mVAR) Models, both as stationary models (\code{mgm()} and  \code{mvar()}) and time-varying models (\code{tvmgm()} and \code{tvmvar()}). In addition, we provide the S3 methods \code{print()} to summarize model objects and \code{predict()} to compute predictions and nodewise errors from all types of models, and the function \code{resample()} to determine the stability of estimates via resampling. Furthermore, \pkg{mgm} provides functions to sample from all four models in full flexibility in order to enable the user to investigate the performance of the estimation algorithms in a particular situation. The output of all estimation functions is designed to allow a seamless visualization with the  \pkg{qgraph} package \citep{qgraph_pkg} and we therefore do not provide our own plotting functions.

\subsection{Related implementations}

Several packages are available to estimate Gaussian Graphical Models (GGMs): the \proglang{R}-packages \pkg{glasso} \citep{friedman_glasso:_2014} and \pkg{huge} \citep{huge_pkg, zhao_huge_2012} implement the graphical lasso \citep{Banerjee_model_2008,friedman_sparse_2008} which maximizes a $\ell_1$-penalized Gaussian log-likelihood. The \pkg{huge} package also allows to estimate GGMs via neighborhood selection \citep{meinshausen_high-dimensional_2006}, in which the neighborhood of each node is estimated separately and then the local estimates are combined to obtain the (global) graphical model. The \proglang{R}-package \pkg{IsingFit} \citep{borkulo_isingfit:_2014, van_borkulo_new_2014} implements a neighborhood selection based method to estimate the binary-valued Ising model \citep[see e.g.][]{wainwright_graphical_2008, Ravikumar_high-dimensional_2010}. The \pkg{XMRF} package \citep{XMRFpackage} allows to estimate Markov Random fields of the Multivariate Gaussian distribution, Ising models, log-linear Poisson based graphical model, regular Poisson graphical models, truncated Poisson graphical models and sublinear Poisson graphical models \citep{yang2015graphical, yang2013poisson}. 

For VAR models, the \pkg{vars} package \cite{varsPackage} allows to fit VAR models with Gaussian noise. The \pkg{BigVAR} package \citep{BigVAR} allows to fit VAR models and structured VAR models with Gaussian noise with structured $\ell_1$- penalties. The \pkg{mlVAR} package \citep{mlVAR} implements multilevel VAR models with Gaussian noise. Graphical VAR models \citep{wild2010graphical}, in which lagged coefficients and contemporaneous effects are estimated simultaneously, can be estimated with the \pkg{graphicalVAR} package \citep{graphicalVAR}.

For time-varying graphical models, there is a \proglang{Python} implementation of the SINGLE algorithm of \cite{monti_estimating_2014} for time-varying Gaussian graphical models \citep{monti_2014} and \pkg{GraphTime} \citep{GraphTime}, a \proglang{Python} implementation of time-varying (dynamic) graphical models based on the (group) fused-lasso as presented by \cite{gibberd2017regularized}. The \proglang{R} package \pkg{tvReg} allows to estimate linear VAR models using kernel smoothing \citep{tvReg}.

\pkg{mgm} goes beyond the above mentioned packages in that it allows to estimate $k$-order MGMs and mVAR models (with any set of lags), compute predictions from them and assess model stability via resampling, while the above packages only allow to do this for special cases. In addition, the output of \pkg{mgm} is designed to allow a seamless visualization of estimated models using the R-package \pkg{qgraph} \citep{qgraph_pkg}. Finally, \pkg{mgm} is the first package that allows to estimate time-varying MGMs and mVAR models.

\subsection{Overview of the paper}

In Section \ref{theory}, we introduce Mixed Graphical Models (MGMs) (Section \ref{theory_mgm}) and mixed Vector Autoregressive (mVAR) models (Section \ref{theory_mvar}), and discuss how to estimate these models in their stationary (Section \ref{theory_mgm_estimation}) and time-varying (Section \ref{theory_timevarying}) versions. In Section \ref{usage}, we illustrate how to use the \pkg{mgm} package to estimate parameters, compute predictions from and visualize stationary MGMs (Section \ref{sec_mgm}), stationary mVAR models (Section \ref{sec_mvar}), time-varying MGMs (Section \ref{sec_tvmgm}) and time-varying mVAR models (Section \ref{sec_tvmvar}). All presented examples are fully reproducible, with code either shown in the paper or provided in the online supplementary material.

\newpage

\section{Background}\label{theory}

In this section we provide basic concepts related to graphical models (Section \ref{theory_gm}), introduce the model classes Mixed Graphical Models (MGMs) (Section \ref{theory_mgm}) and mixed Vector Autoregressive (mVAR) models (Section \ref{theory_mvar}), and show how to estimate these models in their stationary (Section \ref{theory_mgm_estimation}) and time-varying (Section \ref{theory_timevarying}) versions.

\subsection{Graphical Models}\label{theory_gm}

Undirected graphical models are families of probability distributions that respect a set of conditional independence statements represented in an undirected graph $G$ \citep{Lauritzen_graphical_1996}. This connection between probability distribution and graph $G$ is formalized by the \textit{Global Markov Property}, which we define after introducing some notation.

An undirected graph $G = (V,E)$ consists of a collection of nodes $V = \{1, 2, \dots, p\}$ and a collection of edges $E \subset V \times V$. A subset of nodes $U$ is a \textit{node cutset} whenever its removal breaks the graph in two or more nonempty subsets, which is equivalent to $U$ being the set such that all paths from disjoint node sets $S$ and $Q$ go through $U$ \citep{Lauritzen_graphical_1996}. A \textit{clique} is a subset $C \subseteq V$ such that $(s,t) \in E$ for all $s,t \in C$ where $s \neq t$, and is called a \emph{maximal clique} if inclusion of any other node would make it not a clique. The neighborhood $N(s)$ of a node $s \in V$ is the set of nodes that are connected to $s$ by an edge, $N(s) := \{t \in V | (s,t) \in E \}$. Throughout the paper we use the shorthand $X_{\setminus s}$ for $X_{V \setminus \{s\}}$.

To each node $s$ in graph $G$ we associate a random variable $X_{s}$ taking values in a space $\mathcal{X}_s$. For any subset $A \subseteq V$, we use the shorthand $X_{A} := \{X_{s}, s \in A\}$. For three disjoint subsets of nodes, $A$, $B$, and $U$, we write $X_{A} \independent X_{B} | X_{U}$ to indicate that the random vector $X_{A}$ is independent of $X_{B}$ when conditioning on $X_{U}$. We can now define graphical models in terms of the Markov property \citep[e.g.][]{loh2012structure}:


\newtheorem{mrf1}{Definition}\label{definition_1}
\begin{mrf1}
	(Global Markov property). If $X_{A} \independent X_{B} | X_{U}$ whenever $U$ is a node cutset that breaks the graph into disjoint subsets $A$ and $B$, then the random vector $X := ( X_{1}, \dots , X_{p})$ is Markov with respect to the graph $G$.
\end{mrf1}

Note that the neighborhood set $N(s)$ is always a node cutset for $A=\{s\}$ and $B= V \setminus \{s \cup N(s)\}$. 

In the remainder of this paper we focus on exponential family distributions, which are strictly positive distributions. For these distributions the Global Markov property is equivalent to the \textit{Markov factorization property} by the Hammersley-Clifford Theorem \citep{Lauritzen_graphical_1996}. Consider for each clique $C$ in the set of all clique sets $\mathcal{C}$ a clique-compatibility function $\psi_{C}(X_{C})$ that maps configurations $x_C = \{x_s, s \in C \} $ to $\mathbb{R^+}$ such that $\psi_{C}$ only depends on the variables $X_{C}$ corresponding to the clique $C$.

\newtheorem{mrf2}{Definition}\label{definition_2}
\begin{mrf1}
	(Markov factorization property). The distribution of $X$ factorizes according to $G$ if it can be represented as a product of clique functions
	\begin{equation}\label{eq:fac}
	P(X) \propto 
	\prod_{C \in \mathcal{C}} \psi_{C}(X_{C}).
	\end{equation}
	\normalsize
\end{mrf1}

This equivalence implies that if we have distributions that are represented as a product of clique functions, then we can represent the conditional dependence statements in this distribution in a graph $G$. This is the case for the exponential family distributions we use in the present paper

\begin{equation}\label{eq:expfam}
P(X) = 
\exp \left \{ \sum_{C \in \mathcal{C}} \theta_{C} \phi_{C}(X_{C}) - \Phi(\theta)
\right \},
\end{equation}
\normalsize

\noindent
where the functions $\phi_{C}(X_{C}) = \log \psi_{C}(X_{C})$ are sufficient statistic functions specified by the exponential family member at hand (e.g. Gaussian, Exponential, Poisson, etc.),  $\theta_{C}$ are parameters associated with the clique functions and $\Phi(\theta)$ is the log-normalizing constant

\begin{equation*}
\Phi(\theta) = \log \int_\mathcal{X}  \sum_{C \in \mathcal{C}} \theta_{C} \phi_{C}(X_{C}) \nu (dx),
\end{equation*}
\normalsize

\noindent
where depending on the distribution of $X$, the measure $\nu$ is a counting measure or Lesbesgue measure \citep[for details see][]{wainwright2008graphical}.

The graph $G$ represents a \textit{family} of distributions because its edges do not indicate the strength of the dependency and the nodes can represent different conditional distributions. Hence there is a one to one mapping from the density to the graph, but a one to many mapping from graph to densities.

\subsection{Mixed Graphical Models}\label{theory_mgm}

In this section we first introduce the general class of Mixed Graphical Models (\ref{theory_mgm_general}), and then provide the Ising-Gaussian model as a specific example (\ref{theory_mgm_isinggaussian}).

\subsubsection{General Mixed Graphical Models}\label{theory_mgm_general}

In this section, we introduce the class of Mixed Graphical Models (MGMs), which are a special case of the distribution in Equation~\ref{eq:expfam} that  allow one to combine an arbitrary set of conditional univariate members of the exponential family in a joint distribution \citep{yang2014mixed, chen2015selection}.

Consider a $p$-dimensional random vector $X = (X_{1}, \dots,  X_{p})$ with each variable $X_{s}$ taking values in a potentially different set $\mathcal{X}_{s}$, and let $G = (V,E)$ be an undirected graph over $p$ nodes corresponding to the $p$ variables. Now suppose the node-conditional distribution of node $X_{s}$ conditioned on all other variables $X_{\setminus s}$ is given by an arbitrary univariate exponential family distribution

\begin{equation}\label{yang:mixed_gcond}
P(X_{s}|X_{\setminus s}) = 
\exp  \{ 
E_{s}(X_{\setminus s}) \phi_{s}(X_{s}) 
+ B_{s}(X_{s}) - \Phi (X_{\setminus s})  
\},
\end{equation}

\noindent
where the functions of the sufficient statistic $\phi_{s}(\cdot)$ and the base measure $B_{s}(\cdot)$ are specified by the choice of exponential family and the canonical parameter $E_{s}(X_{\setminus s})$ is a function of all variables except $X_{s}$. \cite{wainwright2008graphical} make these functions explicit for a number of exponential family distributions.

These node-conditional distributions are consistent with a joint distribution over the random vector $X$ as in (\ref{eq:fac}), that is Markov with respect to graph $G=(V,E)$ with the set of cliques $\mathcal{C}_k$ of size at most $k$, if and only if the canonical parameters $\{E_{s} (\cdot)\}_{s\in V}$ are a linear combination of products of univariate sufficient statistic functions $\{\phi(X_{r}) \}_{r \in N(s)}$ of order up to $k$

\begin{equation}\label{eq:nodecond.satisfy}
\theta_{s} + \sum_{r \in N(s)} \theta_{s,r} \phi_{r}(X_{r}) + ...
+ \sum_{r_{1}, ..., r_{k-1} \in N(s)} \theta_{r_{1}, ..., r_{k-1}} \prod_{j=1}^{k-1} \phi_{r_{j}} (X_{r_{j}}),
\end{equation}
\normalsize

\noindent
where $\theta_{s\cdot} := \{\theta_{s}, \theta_{s,r}, ..., \theta_{s r_{2}... r_{k}}   \}$ is a set of parameters and $N(s)$ is the set of neighbors of node $s$ according to graph $G$ \citep{yang2014mixed}. Factorizing $p$ conditional distributions as in Equation~\ref{yang:mixed_gcond} gives the joint distribution

\begin{equation}\label{eq:mixed.joint.full}
\begin{split}
P(X) = \exp& \left \{ 
\sum_{s \in V} \theta_{s} \phi_{s} (X_{s}) + \sum_{s \in V} \sum_{r \in N(s)} \theta_{s,r} \phi_{s}(X_{s}) \phi_{r}(X_{r})
+
\right. \\ & \left. 
\dots +  \sum_{r_{1}, ..., r_{k} \in \mathcal{C}} \theta_{r_{1}, ..., r_{k}} \prod_{j=1}^{k} \phi_{r_{j}} (X_{r_{j}}) 
+
\sum_{s \in V} B_{s}(X_{s}) - \Phi(\theta) \right \},
\end{split}
\end{equation}
\normalsize

\noindent
where $\Phi(\theta)$ is the log-normalization constant. 

The dimensionality of the parameter vector $\theta$ depends both on the type of modeled variables and the order of interactions. If one only models continuous variables with pairwise interactions ($k=2$), the MGM simplifies to the multivariate Gaussian distribution which is parameterized by a $1 \times p$ vector of intercepts and a $p \times p$ matrix of ${p}\choose{2}$ partial correlations. Including all 3-way interactions would lead to an additional ${p}\choose{3}$ parameters, etc. At the end of  Section \ref{theory_mgm}, we discuss the dimensionality of the parameter vector in the presence of categorical variables.

Necessary conditions for the mixed density in Equation~\ref{eq:mixed.joint.full} to be normalizable are discussed in \cite{yang2014mixed}. \cite{chen2015selection} show constraints on the parameter space to ensure normalizability for a number of MGMs with at most pairwise interactions. \pkg{mgm} does not allow one to implement the constraints, since the underlying \pkg{glmnet} package does not support the specification of these constraints. However, \cite{trip2018parallel} recently proposed an algorithm that allows to estimate pairwise MGMs with these constraints.

\subsubsection{Example: The Ising-Gaussian Model}\label{theory_mgm_isinggaussian}

We take the Ising-Gaussian model as a specific example of the joint distribution in Equation~\ref{eq:mixed.joint.full}. Consider a random vector $X := (Y, Z)$, where $Y = \{Y_{1}, \dots, Y_{p_1}\}$ are univariate Gaussian random variables, $Z = \{Z_{1}, \dots, Z_{p_2}\}$ are univariate Bernoulli random variables and we only consider pairwise interactions between sufficient statistics. For the univariate Gaussian distribution (with known variance $\sigma^2$) the sufficient statistic function is $\phi_{Y}(Y_{s}) = \frac{Y_{s}}{\sigma_{s}}$ and the base measure is $B_{Y}(Y_{s}) = - \frac{Y_{s}^2}{2\sigma_{s}^{2}}$. The Bernoulli distribution has the sufficient statistic function $\phi_{Z_{s}} = Z_{s}$ and the base measure $B_{Z}(Z_{s}) = 0$. From the MGM joint distribution in Equation~\ref{eq:mixed.joint.full} follows that this mixed density is given by

\begin{equation}\label{eq:mixed.joint.pair.isinggaussian}
\begin{split}
P(Y,Z) \propto \exp & \left \{
\sum_{s \in V_{Y}} \frac{\theta_{s}}{\sigma_{s}} Y_{s} + 
\sum_{r \in V_{Z}} \theta_{r} Z_{r} + 
\sum_{(s,r) \in E_{Y}} \frac{\theta_{s,r}}{ \sigma_{s} \sigma_{r}} Y_{s} Y_{r} + 
\right. \\ & \left.
\sum_{(s,r) \in E_{Z}} \theta_{s,r} Z_{s} Z_{r} +  
\sum_{(s,r) \in E_{YZ}} \frac{\theta_{s,r}}{ \sigma_{s}} Y_{s} Z_{r} - 
\sum_{s \in V_{Y}} \frac{Y_{s}^{2}}{ 2 \sigma^{2}_{s}}
\right \}
,
\end{split}
\end{equation}
\normalsize

\noindent
where the first two terms are thresholds for Gaussian and Bernoulli variables, the third term represents pairwise interactions between Gaussians, the fourth term represents pairwise interactions between Bernoulli variables, the fifth term represents pairwise interactions between Gaussians and Bernoulli variables, and the last term sums over the base measures for the Gaussians.

When the conditional distribution is a Bernoulli random variable $Z_r$, it is given by

\begin{equation}\label{yang_mixed_isinggauss_cising}
\begin{split}
P(Z_{r}|Z_{\setminus r}, Y) \propto \exp &
\left  \{   
\theta_{r} Z_{r} +
\sum_{s \in N(r)_{Z}} \theta_{s,r} Z_{s} Z_{r} +   
\sum_{s \in N(r)_{Y}} \frac{\theta_{s,r}}{\sigma_{r}} Z_{r} Y_{s}  
\right \}.
\end{split}
\end{equation}
\normalsize

\noindent
Note that the conditional distribution in Equation~\ref{yang_mixed_isinggauss_cising} has the same form as the distribution of a single variable conditioned on all remaining variables in an Ising model plus one additional term for interactions between Bernoulli and Gaussian random variables.

When the conditional distribution is a Gaussian random variable $Y_s$, it is given by

\begin{equation*}\label{yang_mixed_isinggauss_cgauss}
P(Y_{s}|Y_{\setminus s}, Z) \propto \exp
\left \{
\frac{\theta_{s}}{\sigma_{s}} Y_{s} +     
\sum_{r \in N(s)_{Y}} \frac{\theta_{s,r}}{\sigma_{s}\sigma_{r}} Y_{s} Y_{r} 
+   
\sum_{r \in N(s)_{Z}} \frac{\theta_{s,r}}{\sigma_{s}} Y_{s} Z_{r} -
\frac{Y_{s}^{2}}{2 \sigma_{s}^{2}} 
\right \}.
\end{equation*}
\normalsize

\noindent
Now, let $\sigma = 1$, factor out $Y_{s}$ and let $\mu_{s} =  \theta_{s} + \sum_{r \in N(s)_{Y}} \theta_{s,r} Y_{r} +  \sum_{r \in N(s)_{Z}} \theta_{s,r} Z_{r}$. Finally, when taking $\frac{\mu_{s}^2}{2}$ out of the log normalization constant, we arrive with basic algebra at the well-known form of the univariate Gaussian distribution with unit variance

\begin{equation*}
P(Y_{s} | Y_{\setminus s}, Z) = \frac{1}{\sqrt{2 \pi}} \exp 
\left \{  
- \frac{(Y_{s} - \mu_{s})^{2}}{2}
\right \}.
\end{equation*}
\normalsize

\subsubsection{Relationship between model parameters and edges in graph}\label{theory_mgm_mapping}

For pairwise MGMs (size of cliques is at most $k = 2$), a pairwise interaction between two continuous variables $s$ and $r$ is parameterized by a single parameter $\theta_{s,r}$. Now, whether the edge between $s$ and $r$ is present depends on whether $\theta_{s,r}$ is zero or not, that is, $(s, r) \in E \iff \theta_{s, r} \neq 0$. Thus, if only pairwise interactions between continuous variables are modeled, any given edge is a function of a \emph{single} parameter. This implies that a \emph{weighted} graph fully represents the parameterization of interactions in the underlying model (or the full parameterization minus the threshold parameters). Interactions between categorical variables with $m > 2$, however, are specified by more than one parameter.  For instance, a pairwise interaction between two categorical variables with $m$ and $u$ categories is parameterized by $R = (m-1) \times (u-1)$ parameters associated with corresponding indicator functions for all $R$ states \citep[e.g.][]{agresti2003categorical}. A pairwise interaction between a categorical variable with $m$ categories and a continuous variable has $R = 1 \times (m-1)$ parameters associated with $m - 1$ indicator functions multiplied with the continuous variable. In this case, $\theta_{s,r}^z$ is a parameter defining the interaction between the nodes $s$ and $t$ indexed by $z \in \{1, \dots, R\}$. In such a situation, an edge is present between $s$ and $r$ if all parameters do \emph{not} have the same value, indicating that not all states have the same probability. In \pkg{mgm} we use the parameterization for multinomial regression of \pkg{glmnet}, which models the probability of each state of the predicted variable, and codes the first category of the predictor variable as the reference category that is absorbed in the intercept \citep[for details see ][]{friedman2010regularization}. This results in $m \times (u-1)$ parameters, where $m$ indicates the number of categories of the predicted variable. In this parameterization, an edge is present if \emph{any} of the parameters in $\theta_{s,r}$ are nonzero, that is, $(s, r) \in E \iff \exists r : |\theta_{s, r}^z| > 0$. Therefore, depending on which variables an edge connects it is defined with respect to one or several parameters.

For general $k$-order MGMs, an edge between nodes $s$ and $r$ is a function of all cliques of size up to $k$ that include both $s$ and $r$. Therefore, for instance, it is not clear from the graph $G$ whether the edge $(s, r)$ is due to a pairwise interaction or from higher order interactions (cliques) that include $s$ and $r$, or both. The number of parameters associated to each clique discussed above for pairwise interactions extends to $k$-order interactions. An interaction between $k$ continuous variables is parameterized by a single parameter $\theta_{r_1, \dots, r_{k-1}}$ and an interaction between $k$ categorical variables is parameterized by $(m_1 - 1) \times \dots \times (m_k - 1)$ parameters, where $m_1, m_2, \dots, m_k$ are the number of categories of each categorical variable.

In this paper we focus mainly on the estimation of pairwise MGMs, where each edge is a function of the parameter(s) of a single pairwise interaction. However, in Section \ref{sec_mgm} we estimate a higher order MGM and visualize the dependency structure in a factor graph. The factor graph representation has the advantage one can still see on which set of cliques a dependency between two nodes depends \citep{koller2009probabilistic}.
{\tiny }

\subsection{Estimating Mixed Graphical Models}\label{theory_mgm_estimation}

In this section, we discuss how to estimate the parameters of a joint distribution of the form as in Equation~\ref{eq:mixed.joint.full} from observations. The graphical model $G$ is then obtained from the parameter estimates as discussed in the previous section.

We know that the joint distribution in Equation~\ref{eq:mixed.joint.full} can be represented as a factorization of univariate conditional distributions. Thus, if we estimate the $p$ univariate conditional distributions with the parameterization in Equation~\ref{eq:nodecond.satisfy}, we obtain the joint distribution. Since all univariate conditional distributions are members of the exponential family, it is possible to estimate the joint distribution in Equation~\ref{eq:mixed.joint.full} by a series of $p$ regressions in the Generalized Linear Model (GLM) framework \cite[see e.g.][]{nelder1972generalized}. From a graphical models perspective this means that we estimate the neighborhood $N(s)$ of each node $s \in V$ and then combine all neighborhoods to obtain an estimate of the graph $G$ \citep{meinshausen2006high}.

In order to obtain parameter estimates that are exactly zero (and thereby imply absent edges in the graph) we minimize the negative log-likelihood $\mathcal{L} (\theta, X)$  together with the $\ell_1$-norm of the parameter vector

\begin{equation}\label{lasso_lossfunction}
\hat{\theta}	= \arg \min_{\theta}  \left \{  \mathcal{L} (\theta, X) + \lambda_n || \theta ||_1  \right \}
,
\end{equation}

\noindent
where $|| \theta ||_1 = \sum_{j=1}^J |\theta_j|$ and $J$ is the length of the parameter vector $\theta$. The negative log-likelihood $ \mathcal{L} (\theta, X)$ is defined by the exponential family distribution of the node at hand. In the Gaussian case, minimizing the negative log-likelihood is equivalent to minimizing the squared loss $  \mathcal{L} (\theta, X) = || X_s - X_{\setminus s}  \theta ||_2^2 $. In other words, we are performing an $\ell_1$-penalized (LASSO) regression in the GLM framework with a link-function appropriate for the node at hand \cite[see e.g.][]{nelder1972generalized}. The $\ell_1$-penalty ensures that the model is identified in the high-dimensional setting $p > n$, where we have more parameters than observations \citep{hastie2015statistical}.

The design matrix is defined with respect to the conditional distribution of node $s$ in the $k$-order MGM. For example, if $k = 2$, the design matrix for the regression on node $s$ contains all other variables or the corresponding indicator functions (for categorical variables). If $k = 3$, the design matrix for the regression on node $s$ contains all other variables or the corresponding indicator functions, plus the products of all pairs of variables in $V_{\setminus s}$, or the $(m - 1) \times (u - 1)$ indicator functions in the case of categorical variables with $m$ and $u$ categories.

To give non-asymptotic guarantees of false and true positive rates for the $\ell_1$-regularized regression estimator it is necessary to put a lower bound $\tau_n$ on the size of the parameters in the true model. This assumption is often called the \emph{beta min condition} \citep[see e.g.][]{hastie2015statistical}. By thresholding estimates at $\tau_n$, we approximately enforce this condition \citep[see also][]{loh2012structure}. For estimating the joint distribution in Equation~\ref{eq:mixed.joint.full} we show in \cite{haslbeck2017mi} that

\begin{equation}\label{tau_threshold}
\tau_n \asymp s_0 \sqrt{\log \frac{p}{n} } \leq s_0 \lambda_n 
,
\end{equation}

\noindent
where $s_0$ is the true number of neighbors. If all variables are continuous, the number of neighbors is equal to the number of nonzero parameter estimates $s_0 = || \theta^* ||_0$, where $\theta^*$ is the true parameter vector. In the case of categorical variables, interactions are parameterized by several parameters. In this case the categorical neighbor is present if at least one of the parameters defining the interaction is nonzero. Since the true parameter vector $\theta^*$ is unknown, we plug in the estimated parameter vector $\hat \theta$ to obtain the \emph{estimated} number of neighbors $ \hat s_0 = || \hat \theta ||_0$. For interactions involving more than one parameter, we plug in the aggregated parameter (see Algorithm \ref{alg_1}). Note that \pkg{mgm} allows to switch off this thresholding (see Section \ref{sec_mgm}). Of course, switching off the thresholding gives a solution that does not have the guarantees of false and true positive rates.

We determine whether an edge is present or not as described in Section \ref{theory_mgm}. In addition, we compute a \emph{weight} from the set of parameters of parameters of each interaction. If the interaction only involves continuous variables there is only one parameter and we take its value. If the interaction involves categorical variables, we take the mean of the absolute value of all parameters as the weight of the edge. From the nodewise regressions we obtain $k$ edge-weights for each $k$-order interaction. For example, for a a pairwise interaction  ($k = 2$) between nodes $s$ and $r$, we obtain one parameter $\theta_{s,r}$ from the regression on $s$ and $\theta_{r,s}$ from the regression on $r$. To obtain a final conditional dependence graph $G$ we need to combine these into a final weight. This can be done either by using the OR-rule (take arithmetic mean of $k$ parameter estimates) or the AND-rule (take arithmetic mean of $k$ parameter estimates if all parameter estimates are nonzero, otherwise set the parameter to zero). Algorithm \ref{alg_1} summarizes this procedure:


\newtheorem{a1}{Algorithm}
\begin{a1}
	\label{alg_1}
	(Estimating Mixed Graphical Models via Neighborhood Regression)
	\begin{enumerate}
		\item For each $s \in V$
		\begin{enumerate}
			\item Construct design matrix defined by $k$, the order of the MGM
			\item Solve the lasso problem in Equation~\ref{lasso_lossfunction} with regularization parameter $\lambda_n$
			\item Threshold the estimates at $\tau_n$
			\item Aggregate interactions with several parameters into a single edge-weight
		\end{enumerate}
		\item Combine the edge-weights with the AND- or OR-rule
		\item Define $G$ based on the zero/nonzero pattern in the combined parameter vector
	\end{enumerate}
\end{a1}

The regularization parameter $\lambda_n$ can be selected using cross-validation or a model-selection criterion such as the Extended Bayesian Information Criterion (EBIC):

\begin{equation}
EBIC_{\gamma}(\hat \theta) =   - 2 \mathcal{L}(\hat \theta) + \hat s_0 \log n + 2 \gamma \hat s_0 \log p
,
\end{equation}

\noindent
where $\mathcal{L}$ is the log likelihood of the conditional density specified by the estimated parameter vector $\hat \theta$, $ \hat s_0$ is the number of nonzero neighbors in the candidate model, and $\gamma$ is a tuning parameter. Note that if $\gamma=0$ the EBIC is equal to the BIC \citep{schwarz1978estimating}. The EBIC has been shown to perform well asymptotically in selecting sparse graphs \citep{foygel2010extended, foygel_high-dimensional_2014} for any value of $\gamma$. In practice, the choice of $\gamma$ will control the trade-off between sensitivity and precision. \cite{foygel2010extended} used values $\gamma \in \{0, .25, .5, .75, 1 \}$ and showed that increasing $\gamma$ from $0$ to $0.25$ led to a considerable decrease in false positives, without increasing false negatives too much. We therefore adopted $\gamma = 0.25$ as a default value. However, to make an \emph{optimal} choice for $\gamma$, it is necessary to take into account the true model, the number of available observations and the cost of false positives and false negatives. While the true model is unknown in real data, a reasonable $\gamma$ can be selected by running a simulation study roughly reflecting the scenario at hand and choosing the $\gamma$ with the most desirable performance. To this end we provide flexible sampling functions (see Section \ref{usage}).

The computational complexity of Algorithm \ref{alg_1} is $\mathcal{O}(p \log (p 2^{k-1}))$. Thus the algorithm does not scale well for large $k$, the order of interactions in the MGM. However, in most situations $k$ will be small, because interactions with a high order are increasingly difficult to interpret and therefore often not of interest.

Note that using a single regularization parameter $\lambda$ for a model including different edge types may lead to a different penalization for different edge types. This is because edge-parameters are scaled with the sufficient statistic they are associated with and this scaling can differ across exponential family members. While we can bring Gaussian variables on the same scale by substracting their mean and dividing by their standard deviation, this is not possible for categorical or Poisson random variables. A potential solution would be to introduce a different penalization parameter for each edge type. But this would make the selection of regularization parameters $\lambda$ considerably more complicated, because now a $u$-dimensional space of $\lambda$ values has to be searched, where $u$ is the number of different edge types. This is why we currently do not have a procedure in \pkg{mgm} that allows different penalties for different edge types.

The performance of Algorithm \ref{alg_1} depends on the number of variables, the order of interactions, type of variables, the size of parameters relative to the variance of associated variables, the sparsity of the parameter vector and the structure of the dependency graph. The best way to determine the performance for a given situation is therefore to obtain it with a simulation study. To this end \pkg{mgm} provides a flexible function to sample from MGMs such that the performance of Algorithm \ref{alg_1} in a given situation can be evaluated via simulations.

\subsection{Mixed Autoregressive Models}\label{theory_mvar}

In Vector Autoregressive (VAR) models, each node $s$ at time point $t$ is modeled as a linear combination of all variables (including $s$) at a set of earlier time points. The standard VAR model is defined with a Gaussian noise process, such that the model can be split up into $p$ conditional Gaussian distributions \citep[see e.g.][]{hamiltontime1994, pfaff_analysis_2008}. Instead of a univariate conditional Gaussian distribution, one can also associate other univariate exponential family members with a given node. This leaves us with an almost identical model and estimation problem as discussed in the previous section (Section \ref{theory_mgm_estimation}). The only difference is that the canonical parameter of the node-conditional at hand is not a function of parameters associated with interactions of variables at the \emph{same time point}, but a function of parameters associated with variables at \emph{previous time points}. To distinguish this VAR model over mixed variables from the VAR model that is typically defined with only Gaussian variables, we call this model \emph{mixed Autoregressive (mVAR)} model.

The mVAR model can be estimated by estimating the parameters of the conditional probability of each variable $s$ as a function of all variables (including itself) at a set of specified previous time points, denoted by $L$. For example, $L = \{1, 2, 3\}$ specifies a VAR model with lags 1, 2 and 3. We introduce a time index as a superscript $t$ for all variables since we are now dealing with time-ordered observations. We then define the canonical parameter $E_{s}^t (X) $ of the conditional distribution $P(X^t_s| X^{t-1}, X^{t-2}, X^{t-3})$ at time $t$ 

\begin{equation}\label{can_par_mVAR}
E_{s}^t (X) = \theta_{s} +  \sum_{j \in L}   \sum_{r \in N(s)} \theta_{s,r}^{t-j} \phi_{r}(X_{r}^{t-j})
.
\end{equation}

We only included pairwise interactions because \pkg{mgm} does not implement higher order interactions for mVAR models. The canonical parameter function in Equation~\ref{can_par_mVAR} defines the negative log-likelihood $\mathcal{L} (\theta, X)$ in Equation~\ref{lasso_lossfunction} and we can therefore use Algorithm \ref{alg_1} with two modifications: first, we define the design matrix as a function of the included lags $L$ instead of the maximal order of the interactions, which we here fix to $k = 2$ (only pairwise interactions). Second, we do not apply an AND/OR rule, because the cross-lagged effect of $X_s^{t-1}$ on $X_r^{t}$ is a different effect than the cross-lagged effect of $X_r^{t-1}$ on $X_s^{t}$ and thus no parameter is estimated twice. Here we state the modified algorithm explicitly:

\newtheorem{a2}{Algorithm}
\begin{a1}
	\label{alg_2}
	(Estimating mixed VAR models via nodewise regression)
	\begin{enumerate}
		\item For each $s \in V$
		\begin{enumerate}
			\item Construct design matrix defined by $L$, the set of included lags
			\item Solve the lasso problem in Equation~\ref{lasso_lossfunction} with regularization parameter $\lambda_n$
			\item Threshold the estimates at $\tau_n$
		\end{enumerate}
		\item Define the directed graphs $D_j$ based on the zero/nonzero pattern in the combined parameter vector for each lag $j \in L$.
	\end{enumerate}
\end{a1} 

The computational complexity of Algorithm \ref{alg_2} is $\mathcal{O}(p \log (p |L|) )$. Similarly to Algorithm \ref{alg_1}, the regularization parameter $\lambda_N$ can be selected using cross validation or an information criterion such as the EBIC.

Note that the directed graphs in the $p \times p \times |L|$ array $D$ are not encoding conditional independence statements as the graph $G$ for MGMs. But they are a useful summary of the parameters of the mixed VAR model, especially because it allows a visualization as a series of directed networks (see Section \ref{sec_mgm} for illustrations).

Note that the performance of Algorithm \ref{alg_2} depends on the number of variables and the number of lags, the type of variables, the size of parameters relative to the variance of associated variables, the sparsity of the parameter vector and the structure of the dependency graph.  \pkg{mgm} offers a flexible function to sample from mixed VAR models such that the performance of Algorithm \ref{alg_2} in a given situation can be evaluated via simulation studies. \cite{haslbeck2017VAR} report the performance of Algorithm \ref{alg_2} in recovering VAR models with Gaussian noise process in a variety of situations.

\subsection{Estimating time-varying models}\label{theory_timevarying}

For both MGMs (Section \ref{theory_mgm}) and mixed VAR models (Section \ref{theory_mvar}) for time series data, we assumed so far that the models are stationary. This means that the observations at each time $t$ point are generated from the same distribution parameterized by $\theta$. In time-varying models we relax this assumption, such that the parameters $\theta^t$ can be different at each time point $t \in \mathcal{T}_n = \{\frac{1}{n}, \frac{2}{n}, \dots , 1 \} $, where $n$ is the number time points in the time series. Note that we use $n$ to denote the number of observations both for cross-sectional data (observations are measurements of different systems from some population) and time series data (repeated measurements of the same system).

Since one cannot estimate a model from a single time point, we have to make assumptions about how the parameters of the true model vary as a function of time. These assumptions are usually assumptions about \emph{local stationarity} \citep[e.g.][]{zhou2010time} and come in one of two flavors: we either assume that there exists a partition $\mathcal{B}$ of $\mathcal{T}_n$ in which time points are consecutive and in each of the subsets $B \in \mathcal{B}$ the model is stationary, that is, $\forall i, j \in B : \theta^i = \theta^j$. These piecewise constant time-varying models can be estimated with a fused lasso penalty, which puts an additional penalty on parameter changes from one time point to the subsequent time point \citep[see e.g.][]{monti_estimating_2014, kolar_estimating_2012, gibberd2015estimating, gibberd2017regularized}. 

The other type of local stationarity, which we focus on in this paper, requires that the model $\theta^t$ is a smooth function of time. In this case we can combine observations close in time for estimation, because we know that their generating models are similar. This idea is implemented by fitting local models $\hat{\theta}^{t^e}$ across the time series, which only give high weight to data points close to the given estimation point $t^e$. The weight function is usually non-negative and symmetric over $t^e$  \citep[see e.g.][]{song_keller:_2009, zhou_time_2010, kolar_estimating_2010, kolar_sparsistent_2009, tao_multiple_2016, chen_inference_2015}. The full time-varying model is then the set of all local estimates $\{ \theta^{e_1}, \theta^{e_2}, \dots, \theta^{|\mathcal{E}|} \}$ at estimation points $\mathcal{E} = \{t^e_1, t^e_2, \dots, t^{|\mathcal{E}|}\} $, where the entries in $\mathcal{E}$ are usually equally spaced across the time series and the number of estimation points $|\mathcal{E}|$ is chosen depending on how fine-grained one would like to describe $\theta^t$ as a function of time $t$.

Stating the above formally, we estimate the model $\theta^{t^e}$ at time point $t^e$ by minimizing a weighted version of the loss function in Equation~\ref{lasso_lossfunction} in Section \ref{sec_mgm_est}

\begin{equation}\label{lasso_lossfunction_weighted}
\hat{\theta}^{t^e}	= \arg \min_{\theta}  \left \{ 
\frac{1}{\sum_{t = 1}^n w_t^{t^e}}
\sum_{t = 1}^n w_t^{t^e} \mathcal{L} (\theta, X^t) + \lambda_n || \theta ||_1  \right \}
,
\end{equation}

\noindent
where $w_t^{t^e}$ is a function of $t$ defined by a kernel centered over $t^e$. Specifically, we define the weight function $w_t^{t^e}$ to be a Gaussian kernel, normalized such that the largest weight is equal to one \citep{zhou2010time}

\begin{equation}\label{kernel_function}
w_t^{t^e} = \frac{Z_t}{\max_{(t \in \mathcal{T}_n )} \left\{\cup_{t} Z_t \right \}},
\;\;\;\; 
\text{where}
\;\;\;\;
Z_t = \frac{1}{\sqrt{2\pi} \sigma} \exp{ \left \{-\frac{(t - t^e)^2}{2\sigma^2} \right \}}
.
\end{equation}

\begin{figure}[h]
	\begin{minipage}[b]{0.47\linewidth}
		\centering
		\includegraphics[width=1\linewidth]{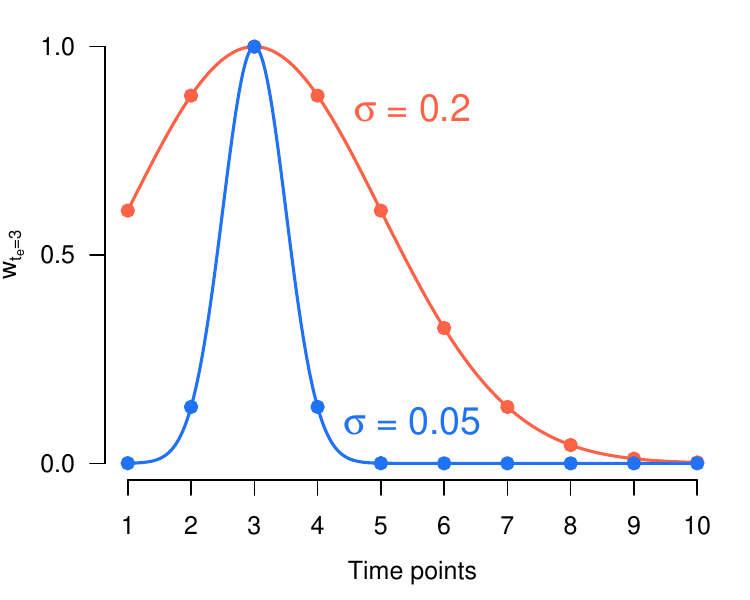}
	\end{minipage}
	\begin{minipage}[b]{0.53\linewidth}
		$$
		\scalemath{.87}{
			\bordermatrix{ \text{Time}   &X_{t, 1} & X_{t-1, 1} & \dots & X_{t, p} & \textcolor{tomato}{w^{ t_{e}=3}} & \textcolor{jssblue}{w^{ t^*_{e}=3}} \cr
				1     & 0.03    &  -0.97  & \dots    & -0.08 & \textcolor{tomato}{0.61} & \textcolor{jssblue}{0.00}    \cr
				2     & 1.15    &  -1.07  & \dots    & -0.56 & \textcolor{tomato}{0.88} & \textcolor{jssblue}{0.14}    \cr
				3     & 0.11    &  0.63   & \dots    & 1.09  & \textcolor{tomato}{1.00} & \textcolor{jssblue}{1.00} \cr
				4     & -1.08   &  0.13   & \dots    & 1.88  & \textcolor{tomato}{0.88} & \textcolor{jssblue}{0.14} \cr
				5     & -0.93   &  1.00   & \dots    & -0.29 & \textcolor{tomato}{0.61} & \textcolor{jssblue}{0.00} \cr
				6     & -1.08   &  0.17   & \dots    & -1.36 & \textcolor{tomato}{0.32} & \textcolor{jssblue}{0.00} \cr
				7     & 0.27    &  -1.72  & \dots    & -1.13 & \textcolor{tomato}{0.14} & \textcolor{jssblue}{0.00} \cr	
				8     & 0.03    &  -1.26  & \dots    & -0.97 & \textcolor{tomato}{0.04} & \textcolor{jssblue}{0.00} \cr
				9     & -1.29   &  -1.05  & \dots    & -0.10 & \textcolor{tomato}{0.01} & \textcolor{jssblue}{0.00} \cr
				10     & -0.07     &  -0.04  & 1.05    & -0.12 & \textcolor{tomato}{0.00} & \textcolor{jssblue}{0.00} }
		}
		$$
		\vspace{.9cm}
	\end{minipage}
	\caption{Illustration of two kernel weighting functions with different bandwidth parameter defined for the estimation point $t^e = 3$; left panel: weights as a function of time; right panel: equivalent representation of the weights across time, combined with the time series data.} \label{tvmodels_figure_general}
\end{figure}

This particular scaling of the weight function has the convenient property that the sum of all weights $n_{\sigma, t^e} = \sum_{t = 1}^n w_t^{t^e}$ (or the area under the curve) used at a given estimation point $t^e$ indicates amount of data used for estimation at $t^e$ relative to the full time series (the full rectangle). Note that we indexed $n_{\sigma, t^e}$ also with the estimation point $t^e$, because less data is used at the beginning and the end of the time series, where the weighting function is truncated (see left panel Figure \ref{tvmodels_figure_general}).

The example in Figure \ref{tvmodels_figure_general} illustrates this estimation procedure. Here we have a time series of $n = 10$ measurements of $p$ continuous variables, and we would like to estimate the model at time point $t^e = 3$. To this end we first define a kernel function $w_t^{t^e}$ as in Equation~\ref{kernel_function}. The bandwidth $\sigma$ of the kernel, which is here equal to the standard deviation of the Gaussian distribution, indicates how many observations close in time we combine to estimate the node at estimation point $t^e$.

Figure \ref{tvmodels_figure_general} displays the kernel function $w_t^{t^e}$ for two different choices of bandwidth, $\sigma = 0.05$ and $\sigma = 0.2$. The kernel function with $\sigma = 0.05$ gives only time points very close to $t^e = 3$ a nonzero weight, while other time points get a weight close to zero and have therefore almost no influence on the parameter estimated at $t^e = 3$. In contrast, the kernel function with $\sigma = 0.2$ distributes weights more evenly, which implies that also time-points quite distant from $t^e = 3$ influence the parameter estimates at $t^e = 3$. The values of both weighting functions at the measured time points are also illustrated together with the data matrix in the right panel of Figure \ref{tvmodels_figure_general}.

The choice of bandwidth involves a trade-off between the sensitivity to time-varying parameters and the stability of the estimates: if we combine only a few observations close in time (small bandwidth $\sigma$) the algorithm can detect parameter-variation at small time scales, however, because we use little data, the estimates will be unstable. If we combine many observations around the estimation point (large $\sigma$), parameter-variation at small time scales will be lost due to aggregation, however, the estimates will be relatively stable. Note that if we keep increasing the bandwidth $\sigma$, the weights on $[0,1]$ will converge to a uniform distribution and give the same estimates as the stationary version of the model, thereby becoming relatively stable, but losing all sensitivity to detect changes in parameters over time.

The ideal bandwidth $\sigma^*$ results in the estimated parameter vector $\hat{\theta}^t$ which minimizes the distance to the true time-varying model $\hat{\theta}^{t*}$ as a function of $\sigma$. We can estimate the ideal bandwidth $\sigma^*$ using a time-stratified cross-validation scheme, where one searches a specified $\sigma$-sequence and selects the $\sigma$ which minimizes the mean (across folds and variables) out of sample prediction error (see Section \ref{sec_tvmgm} for a description of the time-stratified cross-validation scheme).

So far we assumed that the measurements in the time-series are taken at equal time intervals. But this need not be the case, because measurements can be missing randomly or by design. Simply treating the time points as equally distributed leads to an incorrect estimate of the time-varying model. Figure \ref{tvmodels_figure_spacing} illustrates this issue:

\begin{figure}[H]
	\centering
	\includegraphics[width=1\textwidth]{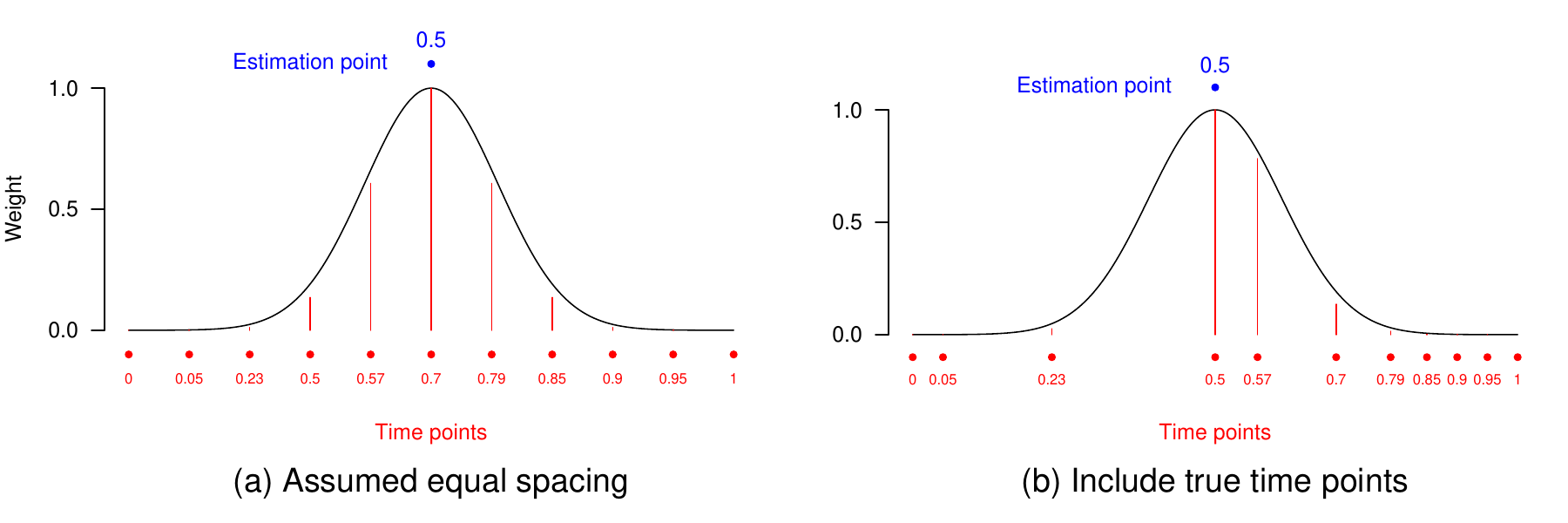}
	\caption{Left panel: the weighting function is computed by assuming that the true time points are equally spaced. Since the true time points are not equally spaced, this creates a mismatch between the time scale of estimation points and the true time scale; right panel: the true time interval is used to compute the weights and hence the two scalings match.} \label{tvmodels_figure_spacing}
\end{figure}

Here we have a time series with $n = 10$ time points, measured at irregular time intervals. In Figure \ref{tvmodels_figure_spacing}(a) we distribute these time points evenly across the time interval, which results in that the assigned time points in the normalized time interval $[0, 1]$ do not correspond to the true time points (values in red). Now if we estimate the time-varying model at time point $t^e = 0.5$, we see that the true time point $0.7$ gets the highest weight. Thus, the model at $t^e = 0.5$ is more strongly influenced by the observations at the true time point $0.7$ than by the observations at time point $0.5$. Clearly, this is undesirable.

In Figure \ref{tvmodels_figure_spacing}(b) we avoid this problem by using the true time points in order to define the weighting function $w_t^{t^e}$. We again estimate the model at $t^e = 0.5$ and see that the time scale of the estimation point is now aligned with the true time scale. This results in a different amount of data used for estimation $n_{\sigma, t^e}$, depending on how many measurements are available around a given time point. If there is less data available, the algorithm becomes more conservative, since we plug in $n_{\sigma, t^e}$ for $n$ in the $\tau_n$ threshold in Equation~\ref{tau_threshold}. In the extreme case where there is no data close to an estimation point, $n_{\sigma, t^e}$ will be extremely small, which implies that the algorithm sets all estimates to zero. This makes sense, because if there is no data available close to a given estimation point $t^e$, we cannot expect to obtain reliable estimates at $t^e$.

Note that the only difference between the stationary models and the time-varying models is that we introduce a weight for each time point in the cost function in Equation~\ref{lasso_lossfunction_weighted} and repeatedly estimate the model at different estimation points. Therefore we can easily adapt the estimation algorithms for the stationary MGM (Algorithm \ref{alg_1}) and mixed VAR model (Algorithm \ref{alg_2}) to their time-varying versions. We first state the algorithm for time-varying MGMs.

\newtheorem{a3}{Algorithm}
\begin{a1}
	\label{alg_3}
	(Estimating time-varying MGMs via kernel-smoothed neighborhood regression)
	\begin{enumerate}
		\item For each estimation point $t^e \in \mathcal{E}$
		\begin{enumerate}
			\item For each variable $s \in V$
			\begin{enumerate}
				\item Construct design matrix defined by $k$, the order of the MGM
				\item Solve the weighted lasso problem in Equation~\ref{lasso_lossfunction_weighted} with regularization parameter $\lambda_n$ and the weighting function $w^{t^e}$ defined by $t^e$ and bandwidth $\sigma$
				\item Threshold the estimates at $\tau_{n_{\sigma, t^e}}$
			\end{enumerate}
			\item Combine the parameter estimates with the AND- or OR-rule
			\item Define $G^e$ based on the zero/nonzero pattern in the combined parameter vector $\theta^e$
		\end{enumerate}
	\end{enumerate}
\end{a1}

Thus we obtain a parameter vector $\theta^{t^e}$ of the MGM in Equation~\ref{eq:mixed.joint.full} and a graph $G^{t^e}$ defined by $\theta^{t^e}$, for each estimation point $t^e \in \mathcal{E}$.  From Algorithm \ref{alg_1} follows that Algorithm \ref{alg_3} has a computational complexity of  $\mathcal{O}(|\mathcal{E}| p \log (p 2^{k-1}) )$.

Similarly, we can adapt Algorithm \ref{alg_2} for the estimation of time-varying mixed VAR models:

\newtheorem{a4}{Algorithm}
\begin{a1}
	\label{alg_4}
	(Estimating time-varying mixed VAR models via kernel-smoothed neighborhood regression)
	\begin{enumerate}
		\item For each estimation point $t^e \in \mathcal{E}$
		\begin{enumerate}
			\item For each variable $s \in V$
			\begin{enumerate}
				\item Construct design matrix defined by the $L$, the set of included lags
				\item Solve the weighted lasso problem in Equation~\ref{lasso_lossfunction_weighted} with regularization parameter $\lambda_n$ and the weighting function $w^{t^e}$ defined by $t^e$ and bandwidth $\sigma$
				\item Threshold the estimates at $\tau_{n_{\sigma, t^e}}$
			\end{enumerate}
			\item Define the directed graphs $D_j^e$ based on the zero/nonzero pattern in the parameter vector $\theta^e$ for each lag $j \in L$.
		\end{enumerate}
	\end{enumerate}
\end{a1}

Here we obtain a parameter vector $\theta^{t^e}$ of the mVAR model and a directed graph $D^{t^e_j}$ for each lag, defined by $\theta^{t^e}$, for each estimation point $t^e \in \mathcal{E}$. From Algorithm \ref{alg_2} follows that Algorithm \ref{alg_4} has a computational complexity of  $\mathcal{O}(|\mathcal{E}| p \log (p |L|) )$.  \cite{haslbeck2017VAR} report the performance of Algorithm \ref{alg_4} in recovering time-varying VAR models with Gaussian noise process for a variety of situations.

Fitting a time-varying model with the above method requires to specify an appropriate bandwidth parameter $\sigma$. In Section \ref{sec_tvmgm}, we describe a time-stratified cross-validation scheme to select $\sigma$ in a data-driven way.  The EBIC is not suitable to select $\sigma$. The reason is that threshold (intercept) parameters are neither included in the $\ell_1$-penalty, nor in the EBIC. This results in the EBIC selecting always the model with the smallest specified bandwidth, which includes no interaction parameters, but achieves an extremely good fit through highly local (time-varying) thresholds (intercepts). This problem is avoided when using a cross-validation scheme, where fitting local means leads to high out-of-fold prediction error. 

Note that the performance of Algorithm \ref{alg_3} and \ref{alg_4} depends on the number of variables, the type of variables, the size of parameters relative to their variance, the sparsity of the parameter vectors, the structure of the dependency graph and the how non-linear the parameters vary as a function of time. The best way to obtain the performance of Algorithm \ref{alg_3} and \ref{alg_4} is to set up a suitable simulation study. To this end \pkg{mgm} offers flexible functions to sample from time-varying MGMs and time-varying mVAR models.


\newpage

\section[UsageAndExamples]{Usage and Examples}\label{usage}

The \pkg{mgm} package can be installed from the Comprehensive R Archive Network (CRAN) (\url{http://CRAN.r-project.org/}):

\begin{Sinput}
  install.packages("mgm")
  library(mgm)
\end{Sinput}

In the following four sections, we show for each of the four model types how to

\begin{enumerate}
	\item sample observations from a specified model
	\item estimate the model from data
	\item make predictions from an estimated model
	\item visualize the estimated model
	\item and assess the stability of estimates.
\end{enumerate}

The sampling functions are included to enable the user to determine the performance of the estimation algorithm in a specific situation via simulations. All used datasets are loaded automatically with the \pkg{mgm}-package. All analyses in the paper are fully reproducible, and the necessary code is either shown in the paper or can be found in the online supplementary material or the Github repository \url{https://github.com/jmbh/mgmDocumentation}. For all code examples we use the \pkg{mgm} version 1.2-7.

\subsection{Stationary Mixed Graphical Models}\label{sec_mgm}

In this section we first use a simulated data set to show how to estimate a pairwise MGM, compute predictions from it, visualize it and assess the stability of its parameters. Then we fit a pairwise MGM to a larger empirical data set related to Autism Spectrum Disorder (ASD). Finally, we give an example of a higher-order MGM by showing how to estimate a $k=3$ MGM to a data set consisting of symptoms of Post-traumatic Stress Disorder (PTSD).

\subsubsection{Estimating Mixed Graphical Models}\label{sec_mgm_est}

In this section we show how to use the function \code{mgm()} to estimate a pairwise MGM to a data set with $n=500$ observations of two continuous, and two categorical with $m = 2$ and $u = 4$ categories, respectively. The true model includes the pairwise interactions 1-4, 2-3 and 1-2. For the exact parameterization of the true model and for a description of how to sample from this MGM using the \code{mgmsampler()} function see the section on sampling below.

Next to the data, we specify the type of each variable (\code{"g"} for Gaussian, \code{"p"} for Poisson, \code{"c"} for categorical) and the number of levels of each variable (1 for continuous variables by convention). Here we use the example data \code{mgm_data} which is automatically loaded with \pkg{mgm}. Next, we indicate the order of the graphical model: we choose $k = 2$, which corresponds to a pairwise MGM (containing at most 2-way interactions). If we specified $k = 3$, we would fit an MGM including all 2-way and all 3-way interactions, $k=4$ would include all 2-way, 3-way and 4-way interactions, etc. After that, we specify how we select the penalization parameter $\lambda_n$ in Algorithm \ref{alg_1}. The two available options are the EBIC or cross-validation. Here we choose cross-validation with 10 folds. If not otherwise specified via the argument \code{lambdaSeq}, the considered $\lambda$-sequence is determined as in the \pkg{glmnet} package: a sequence is defined from $\lambda_{max}$, the smallest (data derived) value for which all coefficients are zero, and $\lambda_{min}$, a fraction of $\lambda_{max}$, which is 0.01 in the high-dimensional setting ($n < p$) and $0.0001$ if $n > p$. Finally, indicate that estimates across neighborhood regressions should be combined with the AND-rule. Since we use cross-validation, we set a random seed outside the function to ensure that the analysis is reproducible.

\begin{Sinput}
  R> set.seed(1)
  R> fit_mgm <- mgm(data = mgm_data$data,
  +                type = c("g", "c", "c", "g"),
  +                levels = c(1, 2, 4, 1), 
  +                k = 2, 
  +                lambdaSel = "CV",
  +                lambdaFolds = 10,
  +                ruleReg = "AND")
\end{Sinput}

\code{mgm()} returns a list with the following entries: \code{fit_mgm$call} returns the call of the function; \code{fit_mgm$pairwise} contains the weighted adjacency matrix and the signs (if defined) of the parameters in the weighted adjacency matrix; \code{fit_mgm$interactions} contains a list that shows all recovered interactions (cliques) and a list that returns the parameters associated with all cliques; \code{fit_mgm$intercepts} stores all estimated thresholds/intercepts and \code{fit_mgm$nodemodels} is a list with the $p$ \pkg{glmnet} objects from which all above results are computed. We inspect the weigthed adjacency matrix stored in \code{fit_mgm$pairwise$wadj}

\begin{Soutput}
  R> round(fit_mgm$pairwise$wadj, 2)
       [,1] [,2] [,3] [,4]
  [1,] 0.00 0.74 0.00 0.47
  [2,] 0.74 0.00 0.14 0.00
  [3,] 0.00 0.14 0.00 0.00
  [4,] 0.47 0.00 0.00 0.00
\end{Soutput}

\noindent
and see that we correctly recovered the pairwise dependencies 1-4, 2-3 and 1-2. The list entry \code{fit_mgm$pairwise$signs} indicates the sign for each interaction, if a sign is defined. By default, a sign is only defined for interactions between non-categorical variables (Gaussian, Poisson). Interactions involving categorical variables with $m > 2$ categories are defined by more than one parameter and hence no sign can be defined. The function \code{showInteraction()} provides an alternative way to inspect a given interaction. For instance, one can obtain the details about the interaction 1-4 like this:

\begin{Soutput}
  R> showInteraction(object = fit_mgm, 
  +                 int = c(1,4))
  Interaction: 1-4 
  Weight:  0.4676953 
  Sign:  1  (Positive)
\end{Soutput}

We use the \pkg{glmnet} package to fit the regularized nodewise regressions, which directly models the probabilities of categorical variables instead of the ratio relative to a reference category. This is possible, because the regularization ensures that this model is identified \citep[for details see ][]{friedman2010regularization}. This means that an interaction between two categorical variables $X_1 \in \{1, \dots, m\}$ and $X_2 \in \{1, \dots, u\}$ has $m \times (u-1)$ parameters in the regression on $X_1$ and $u \times (m-1)$ parameters in the regression on $X_2$. In addition, all estimation functions in \pkg{mgm} also allow an overparameterization (specified via the argument \code{overparameterize = TRUE}), where an indicator function is defined for \emph{each} state of the categorical predictor variable. In the previous example of a pairwise interaction, this leads to $m \times u$ parameters specifying the interaction between $X_1$ and $X_2$. The overparameterization is useful when one is interested in parameters associated with indicator functions that are otherwise absorbed by the threshold (intercept) parameters (also called reference category). We give an example for estimating a $k=3$ order MGM at the end of this section.

If the argument \code{binarySign} is set to \code{TRUE}, all binary variables have to be coded as $\{0,1\}$ and a sign is defined in the following way: for an interaction between two binary variables $X_1, X_2 \in \{0, 1\}$, if the parameter associated with the indicator function $\mathbb{I}_{X_2 = 1}$ in the equation modeling $P(X_1 = 1)$ has a positive sign (which implies that the parameter associated with $\mathbb{I}_{X_2 = 1}$ in the equation modeling $P(X_1 = 0)$ has a negative sign, see \cite{friedman2010regularization}), then we assign a positive sign to the binary-binary interaction. For an interaction between a binary variable $X_1$ and a \emph{continuous} variable $X_2$ we take the sign of the parameter associated with $X_2$ in the equation modeling $P(X_1 = 1)$. In addition, it is possible to specify a weight for each observation via the argument \code{weights} to perform weighted regression.

In the example above we used an $\ell_1$-penalized GLM to estimate the MGM, which implies that we assume that the true MGM is sparse. However, a different penalty may be appropriate in some situations. Via the argument \code{alphaSeq} one can specify any convex combination of the $\ell_1$- and $\ell_2$-penalty (the elastic net penalty, see \cite{zou2005regularization}). \code{alphaSeq = 1} corresponds to the $\ell_1$-penalty (default) and \code{alphaSeq = 0} to the $\ell_2$-penalty. If a sequence of values is provided to \code{alphaSeq}, the function will select the best $\alpha$ value based on the EBIC or cross validation, specified via the argument \code{alphaSel}.

\subsubsection{Making Predictions from Mixed Graphical Models}\label{sec_mgm_pred}

We now use the \code{predict()} function to compute predictions and nodewise errors from the model estimated in the previous section. This function takes the model object and data of the same format as the data used for estimation as input. It also allows to specify which error functions should be used to compute nodewise prediction errors. The error functions $F(\hat{y}, y)$ for continuous and categorical variables are specified via the \code{errorCon} and \code{errorCat} arguments, respectively. Here we specified the Root Mean Squared Error (\code{"RMSE"}) and the proportion of explained variance (\code{"R2"}) as error functions for the continuous variables, and the proportion of correct classification (or accuracy, \code{"CC"}) and the normalized proportion of correct classification (\code{"nCC"}) for categorical variables. \code{"nCC"} indicates the increase in accuracy beyond the intercept model, divided by the maximal possible increase, and thereby captures how well a node is predicted by other nodes beyond the intercept model. Specifically, let $\mathcal{A} = \frac{1}{n} \sum_{i=1}^{n} \mathbb{I} (y_i = \hat{y}_i)$ be the proportion of correct classifications, and let $p_0, p_1, \dots p_m$ be the marginal probabilities of the categories, where $\mathbb{I}$ is the indicator function for the event   $R_i = \hat{R}_i$. In the binary case these are $p_0$ and $p_1 = 1 - p_0$. We then define normalized accuracy as

$$ \mathcal{A} _{\text{norm}}  = \frac{\mathcal{A}  - \max\{p_0, p_1, \dots, p_m\}}{1 - \max\{p_0, p_1, \dots, p_m\}}. $$

For details see \cite{haslbeck2016well}. If one is not interested in computing nodewise prediction errors, the arguments \code{errorCon} and \code{errorCat} can be simply ignored.

We provide the \pkg{mgm} fit object and the data as input arguments and a choice of prediction error measures to the  \code{predict()} function:

\begin{Sinput}
  R> pred_mgm <- predict(object = fit_mgm, 
  +                     data = mgm_data$data,
  +                     errorCon = c("RMSE", "R2"),
  +                     errorCat = c("CC", "nCC"))
\end{Sinput}

The output object \code{pred_mgm} is a list that contains the function call, the predicted values, the predicted probabilities of each category in case the model includes categorical variables, and a table with nodewise prediction errors. Here we print the nodewise error table in the console:

\begin{Soutput}
  R> pred_mgm$errors
       Variable Error.RMSE Error.R2 Error.CC Error.nCC
  [1,]        1      0.781    0.389       NA        NA
  [2,]        2         NA       NA    0.838     0.206
  [3,]        3         NA       NA    0.342     0.000
  [4,]        4      0.854    0.270       NA        NA
\end{Soutput}

The RMSE and $R^2$ are shown for the two continuous variables, the accuracy and normalized accuracy are shown for the two categorical variables. It is possible to provide an arbitrary number of customary error functions for both continuous and categorical variables to \code{predict()}, for details see \code{?predict.mgm}.

In this example we used the same data for estimation and prediction, which means that we computed \emph{within sample} prediction errors. In order to evaluate how well the model generalizes out of sample, the predictions have to be made on a fresh test data set. This can be done by providing new data of the same format to the \code{predict()} function.

\subsubsection{Visualizing Mixed Graphical Models}\label{sec_mgm_viz}

We visualize interaction parameters of the pairwise model together with the nodewise errors using the \pkg{qgraph} package \citep{qgraph_pkg}. To this end we first install and load the \pkg{qgraph} package and compute a vector containing the nodewise errors we would like to display:

\begin{Sinput}
  R> install.packages("qgraph")
  R> library(qgraph)
  R> errors <- c(pred_mgm$errors[1, 3], 
  +             pred_mgm$errors[2:3, 4], pred_mgm$errors[4, 3])
\end{Sinput}

Here we decided to display the proportion of explained variance for the continuous variables and the accuracy for the categorical variables. In order to plot the model, one provides the weighted adjacency matrix and the errors to the function \code{qgraph()}. We also provide a matrix of edge colors that specify the sign of each interaction (green = positive, red = negative, grey = undefined) that is stored in the \pkg{mgm} fit object. Finally we provide colors for the different error measures and variable names for the legend.

\begin{Sinput}
  R> qgraph(fit_mgm$pairwise$wadj, 
  +        edge.color = fit_mgm$pairwise$edgecolor, 
  +        pie = errors, 
  +        pieColor = c("lightblue", "tomato", "tomato", "lightblue"), 
  +        nodeNames = c("Gaussian", "Categorical; m = 2", 
  +                      "Categorical; m = 4", "Gaussian"),
  +        legend = TRUE)
\end{Sinput}

Figure \ref{mgm_p4_example} shows the resulting visualization:

\begin{figure}[H]
	\centering
	\includegraphics[width=.55\textwidth]{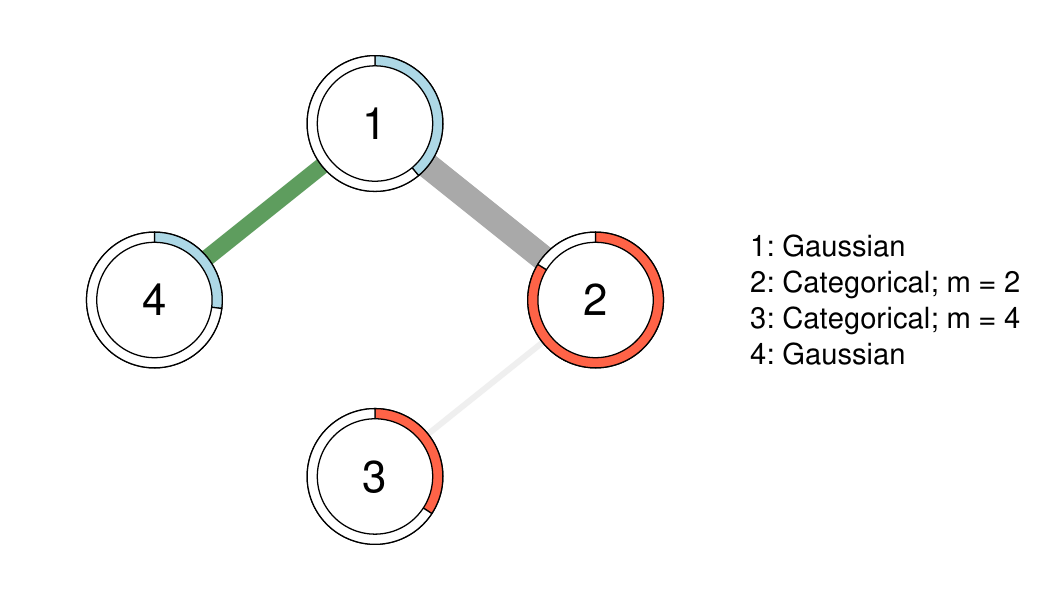}
	\caption{Visualization of the edge-parameters and nodewise errors of the estimated MGM. Green edges indicate positive relationships. Grey edges indicate pairwise interactions for which no sign is defined (interactions involving categorical variables). The width of the edges is proportional to the absolute value of the associated edge-parameter.}\label{mgm_p4_example}
\end{figure}

The green edge between variable 1 and variable 2 indicates a positive linear relationship between the two Gaussian variables and the two grey edges indicate relationships between categorical variables, for which no sign is defined. The exact nature of these interactions can be found by inspecting them using the output object of the \code{showInteraction()} function. The width of the edges is proportional to the size of the corresponding edge-parameter. The blue rings indicate the proportion of variance explained by neighboring nodes for the Gaussian variables, and the red rings indicate the accuracy of the categorical nodes.

\subsubsection{Bootstrap Sampling Distributions}\label{sec_mgm_resample}

Obtaining the sampling distributions for parameter estimates can be useful if one is interested in the stability of estimates \citep{hastie2015statistical}. The function \code{resample()} obtains empirical sampling distributions with the nonparametric bootstrap \citep{efron1992bootstrap, efron1994introduction}. Its input is the \pkg{mgm} model object \code{fit_mgm} (the output of the function \code{mgm()}, see above), the data, and the desired number of bootstrap samples $B$ via the argument \code{nB}. The argument \code{quantiles} specifies lower/upper quantiles of the sampling distributions, which are added to the output. Here we choose \code{quantiles = c(.05, .95)}. Finally, we set a random seed to make the analysis reproducible.

\begin{Sinput}
  R> set.seed(1)
  R> res_obj <- resample(object = fit_mgm, 
  +                     data = mgm_data$data, 
  +                     nB = 50,
  +                     quantiles = c(.05, .95))               
\end{Sinput}

The bootstrapped sampling distributions of the edge weights can be found in a $ B \times p \times p$ array stored in the list entry \code{res_obj$bootParameters}. For example, the vector of length $B$ with the bootstrapped sampling distribution of the weight of the edge 3-4 is stored in the entry \code{res_obj$bootParameters[, 3, 4]}. The output object \code{res_obj} also contains the specified lower/upper quantiles of each sampling distribution, the function call, the $B$ estimated models and the running time for each bootstrap sample. The function \code{plotRes()} provides a plot that summarizes the bootstrapped sampling distributions. For each edge weight, it displays the proportion of nonzero estimates across all $B$ models, printed at the arithmetic mean of the sampling distribution. In addition, it displays specified lower/upper quantiles. Here we choose the 5\% and 95\% quantiles by setting \code{quantiles = c(.05, .95)}. 

\begin{Sinput}
  R> plotRes(object = res_obj, 
  +         quantiles = c(.05, .95))
\end{Sinput}

\begin{figure}[H]
	\centering
	\includegraphics[width=.55\textwidth]{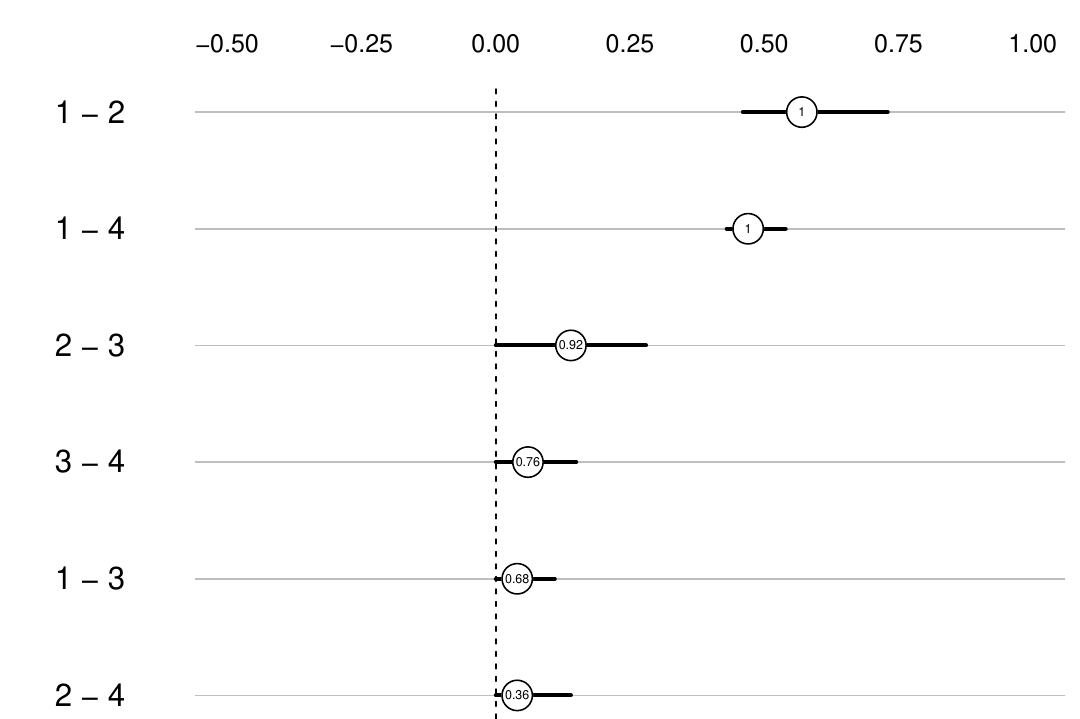}
	\caption{Summaries of bootstrapped sampling distributions separately for the weight of each edge. The value indicates the proportion of nonzero estimates across the $B$ bootstrap samples and is plotted on the airthmetic mean of the sampling distribution. The black horizontal lines indicate the $0.05$ and $0.95$ quantiles of the bootstrapped sampling distribution.}\label{fig_mgm_p4_resample}
\end{figure}

The resulting plot is displayed in Figure \ref{fig_mgm_p4_resample}. It shows that the sampling distributions for the edges 1-2 and 1-4 are located far from zero, have a small standard deviation and 100\% of the $B$ estimates were nonzero. For the edges 3-4, 2-3 and 1-3 that are absent in the true graph, the sampling distribution is close to zero and the proportion of estimated nonzero effects is much smaller.

While bootstrapped sampling distributions are useful to determine the stability of estimates, they are not suited for performing hypothesis tests, for instance, against the null null hypothesis that the population parameter is equal to zero. The reason is that the sampling distributions of parameters obtained with $\ell_1$-regularized regression have a zero mass around zero \citep{buhlmann2014high}. The R-package \emph{bootnet} \citep{epskamp2018estimating} also implements a bootstrap scheme for \pkg{mgm} objects and provides similar plotting options.

\subsubsection{Sampling from Mixed Graphical Models}\label{sec_mgm_sampling}

Here we illustrate how to use the function \code{mgmsampler()} to create the dataset \code{mgm_data} that was used for estimation above. These data were created by specifying an MGM consisting of two continuous-Gaussian nodes (\code{"g"}), and two categorical nodes (\code{"c"}) with $m=2$ and $u=4$ categories, and three pairwise interactions between these four variables. A third option in \code{mgmsampler()} are Poisson nodes (\code{"p"}). Note that we use the overparameterized representation of interactions between categorical variables to specify the model, which means that the pairwise interaction between the categorical variables has $m \times u$ parameters. We begin by specifying the type and number of categories for each node. By convention, for continuous variables we set the number of categories to 1.

\begin{Sinput}
  R> type <- c("g", "c", "c", "g")
  R> level <- c(1, 2, 4, 1)
\end{Sinput}

Next, we specify a list containing the thresholds for each variable:

\begin{Sinput}
  R> thresholds <- list()
  R> thresholds[[1]] <- 0
  R> thresholds[[2]] <- rep(0, level[2])
  R> thresholds[[3]] <- rep(0, level[3])
  R> thresholds[[4]] <- 0
\end{Sinput}

We specify a zero threshold (intercept) for the two Gaussian nodes, and for each of the categories of both categorical variables. Thresholds correspond to the first summation in the joint MGM density in Equation~\ref{eq:mixed.joint.full}. Next, we specify a vector containing the standard deviations for the Gaussian variables:

\begin{Sinput}
  R> sds <- rep(1, 4)
\end{Sinput}

The entries in \code{sds} corresponding to non-Gaussian nodes (here 2 and 3) are ignored. Finally, we specify three pairwise interactions between the variables 1-2, 2-3 and 1-4 in two steps: first, we create a matrix, in which each row indicates one pairwise interaction:

\begin{Sinput}
  R> factors <- list()
  R> factors[[1]] <- matrix(c(1,4,
  +                          2,3,
  +                          1,2), ncol = 2, byrow = T)
\end{Sinput}

We assign the matrix to the first list entry \code{factors[[1]]}, which contains pairwise interactions. The second list entry \code{factors[[2]]} contains a $q \times 3$ matrix of $q$ 3-way interactions, the third entry contains a $w \times 4$ matrix of $w$ 3-way interactions, etc. Since we only specify pairwise interactions in this example, we only use the first entry. A description and examples of how to specify higher order interactions are given in the help file \code{?mgmsampler}. In a second step, we specify the parameters of the three interactions:

\begin{Sinput}
  R> interactions <- list()
  R> interactions[[1]] <- vector("list", length = 3)
  
  R> # 2-way interaction: 1-4
  R> interactions[[1]][[1]] <- array(.5, dim = c(level[1], level[4]))
  
  R> # 2-way interaction: 2-3
  R> int_2 <- matrix(0, nrow = level[2], ncol = level[3])
  R> int_2[1, 1:2] <- 1
  R> interactions[[1]][[2]] <- int_2
  
  R> # 2-way interaction: 1-2
  R> int_1 <- matrix(0, nrow = level[1], ncol = level[2])
  R>int_1[1, 1] <- 1
  R>interactions[[1]][[3]] <- int_1
\end{Sinput}

The interaction between the continuous variables 1-2 is parameterized by one parameter with value $0.5$. The interaction between the two categorical variables is specified by a $2 \times 4$ parameter matrix. We give the entries $(1,1)$ and $(1,2)$ a value of 1, which means that these two states have a higher probability than the remaining states, which are associated with a value of 0. Finally,  we specify the interaction between the continuous-Gaussian node 1 and the binary node 2, which has two parameters associated with the two indicator functions for the binary variable multiplied with the continuous variable. Now we provide these arguments to the \code{mgmsampler()} function, together with $n = 500$, which samples 500 observations from the model:

\begin{Sinput}
  R> set.seed(1)
  R> mgm_data <- mgmsampler(factors = factors,
  +                        interactions = interactions,
  +                        thresholds = thresholds,
  +                        sds = sds,
  +                        type = type,
  +                        level = level,
  +                        N = 500)
\end{Sinput}

The function returns a list containing the function call in \code{mgm_data$call} and the data in \code{mgm_data$data}. For more details on how to specify $k$-order MGMs we refer the reader to the help file \code{?mgmsampler}.

\newpage

\subsubsection{Application: Autism and Well-being}\label{sec_mgm_application}

Here we show how to estimate an MGM on a real data set consisting of responses of 3521 individuals from the Netherlands, who were diagnosed with Autism Spectrum Disorder (ASD), to 28 questions on demographics, psychological aspects, conditions of the social environment and medical measurements \citep[for details see][]{Begeer_Allemaal_2013, deserno2016multicausal}. The dataset is included in the \pkg{mgm} package and loaded automatically. It includes continuous variables, count variables and categorical variables (see \code{autism_data_large$type}), and the latter have between 2 and 5 categories (see \code{autism_data_large$level}))

We choose a pairwise model ($k = 2$) and select the regularization parameters $\lambda_n$ using the EBIC with a hyper-parameter $\gamma = 0.25$:

\begin{Sinput}
  R> fit_ADS <- mgm(data = autism_data_large$data, 
  +                type = autism_data_large$type,
  +                level = autism_data_large$level,
  +                k = 2, 
  +                lambdaSel = "EBIC", 
  +                lambdaGam = 0.25)
\end{Sinput}

The $28 \times 28$ weighted adjacency matrix is too large to be displayed here. Instead, we directly visualize it using the \pkg{qgraph} package. In addition to the weighted adjacency matrix and the matrix containing edge colors that indicate the signs of edge parameters, we also provide a grouping of the variables into the categories \emph{Demographics}, \emph{Psychological}, \emph{Social environment} and \emph{Medical} measurements as well as colors for the grouping, both of which are contained in the data list \code{autism_data_large}. The remaining arguments are chosen to improve the visualization, for details we refer the reader to the help file \code{?qgraph}.

\begin{Sinput}
  R> qgraph(fit_ADS$pairwise$wadj, 
  +        layout = "spring", repulsion = 1.3,
  +        edge.color = fit_ADS$pairwise$edgecolor, 
  +        nodeNames = autism_data_large$colnames,
  +        color = autism_data_large$groups_color, 
  +        groups = autism_data_large$groups_list,
  +        legend.mode="style2", legend.cex=.4, 
  +        vsize = 3.5, esize = 15)
\end{Sinput}

\begin{figure}[h]
	\centering
	\includegraphics[width=.9\textwidth]{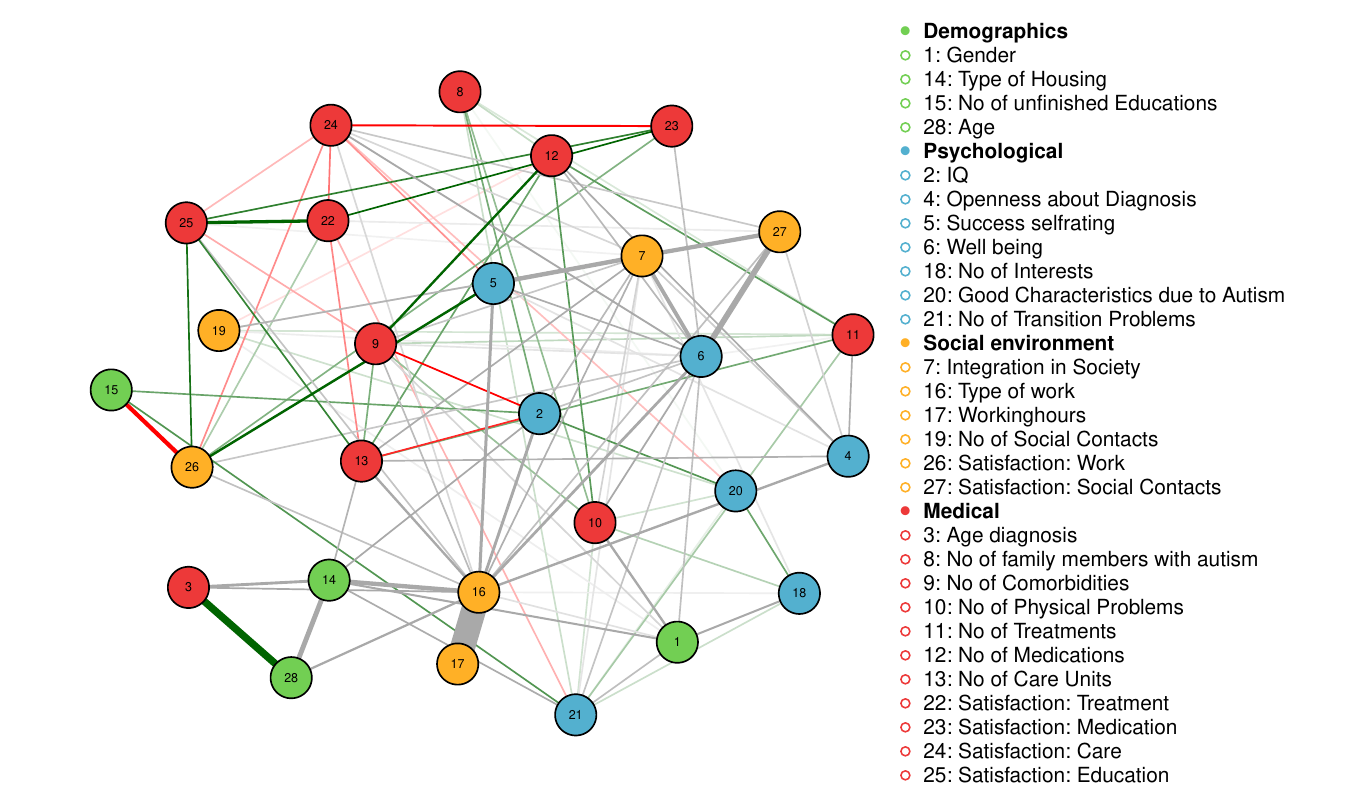}
	
	\caption{Visualization of the MGM estimated on the autism dataset. Green edes indicate positive relationships, red edges indicate negative relationships and grey edges indicate relationships involving categorical variables for which no sign is defined. The width of the edges is proportional to the absolute value of the edge-parameter. The colors of the nodes map to the different domains \emph{Demographics}, \emph{Psychological}, \emph{Social Environment} and \emph{Medical}.}\label{Fig_mgm_application_Autism}
\end{figure}

The resulting visualization is shown in Figure \ref{Fig_mgm_application_Autism}. The layout of node positions was computed with the Fruchterman Reingold algorithm, which places nodes such that all the edges are of more or less equal length and there are as few crossing edges as possible \citep{fruchterman1991graph}. Green edges indicate positive relationships, red edges indicate negative relationships and grey edges indicate relationships involving categorical variables for which no sign is defined. The width of the edges is proportional to the absolute value of the edge-weight. The node color indicates to the different categories of variables.

We observe a strong positive relationship between \emph{age} and \emph{age of diagnosis}, which makes sense because the two variables are logically connected. The negative relationship between \emph{number of unfinished educations} and \emph{satisfaction at work} seems plausible, too. Well-being is strongly connected in the graph, with the strongest connections to \emph{satisfaction with social contacts} and \emph{integration in society}. These three variables are categorical variables with 5, 3 and 3 categories, respectively. In order to investigate the exact nature of these interactions, one can look up all parameters using the function \code{showInteraction()}.

\subsubsection{Estimating higher-order Mixed Graphical Models}\label{sec_mgm_hoi}

In the previous section, we focused on the estimation of pairwise ($k=2$) MGMs. Here, we show how to estimate an MGM of order $k=3$ to a dataset consisting of Post-traumatic Stress Disorder (PTSD) symptoms reported from 344 survivors of the Wenchuan earthquake in China reported in \cite{mcnally2015mental}. The data is loaded automatically with \pkg{mgm} and includes the following symptoms:

\begin{Soutput}
  R> PTSD_data$names
  [1] "intrusion" "dreams"    "flash"     "upset"     "physior"   "avoidth"  
\end{Soutput}

We first specify the data, type, levels and the desired method to select the regularization parameter $\lambda$, similarly to the pairwise MGM. But here we specify with \code{k = 3} to estimate \emph{all} pairwise and \emph{all} 3-way interactions. 

In addition, we choose to use the overparameterized version of the representation of categorical variables by setting \code{overparameterize = TRUE}. This results in that all states of categorical variables up to degree $k$ are modeled explicitly. This overparameterization is possible due to the $\ell_1$-penalization \citep[for details see][]{friedman2010regularization}. The standard and the overparameterized parameterization are statistically equivalent and therefore one has to choose one over the other based on which parameterization lends itself to the most useful interpretation in a given application: if it is more sensible to compare all categories to a reference category the standard parameterization is preferable. If one is interested in all categories equally, the overparameterization might be better. We call \code{mgm()} with the above discussed specifications:

\begin{Sinput}
  R> fit_mgmk <- mgm(data = PTSD_data$data, 
  +                 type = PTSD_data$type, 
  +                 level = PTSD_data$level,
  +                 lambdaSel = "EBIC",
  +                 lambdaGam = 0.25,
  +                 k = 3, 
  +                 overparameterize = TRUE)
\end{Sinput}

The output object \code{fit_mgmk} has the same structure as the pairwise MGM discussed above. We still find the pairwise interactions in \code{fit_mgmk$pairwise} but these do not represent the full parameterization anymore, since we also estimated 3-way interactions. All interactions that have been estimated to be nonzero can be found in the list \code{fit_mgmk$interactions}: the entry \code{fit_mgmk$interactions$indicator} contains a list showing all nonzero estimated interactions, separately for each order (here 2 and 3):

\begin{Soutput}
  R> fit_mgmk$interactions$indicator
  [[1]]
  [,1] [,2]
  [1,]    1    2
  [3,]    4    5
  
  [[2]]
  [,1] [,2] [,3]
  [1,]    1    3    4
  [2,]    1    3    5
  [3,]    2    3    4
  [4,]    3    5    6
  [5,]    4    5    6
\end{Soutput}

The  output indicates that we estimated two nonzero pairwise interactions, and five nonzero 3-way interactions. For example, the third row in the second list entry indicates that there is a 3-way interaction between variables 2-3-4 (\emph{Dreams},  \emph{Flashbacks} and  \emph{Upset}). The list \code{fit_mgmk$interactions} also contains additional entries for the strength of each interaction, and all parameters specifying the interaction (more than one parameter in case of categorical variables, see Section \ref{theory_gm}). 

If we were to visualize the dependency structure of this $k=3$-order MGM in a common undirected graph, we would lose the information about on which interaction(s) a dependency (edge) is based on. For instance, an edge between the nodes 1 and 2 could either be due to a pairwise interaction between 1 and 2, or due to any 3-way interaction including the nodes 1 and 2, or both. A visualization that allows to represent different orders of interactions is the factor graph \citep[e.g.][]{koller2009probabilistic}. A factor graph is a bipartite graph that includes nodes for variables on the one hand, and nodes for interactions on the other hand. We use the function \code{FactorGraph()} to plot such a factor graph

\begin{figure}[h]
	\centering
	\includegraphics[width=1\textwidth]{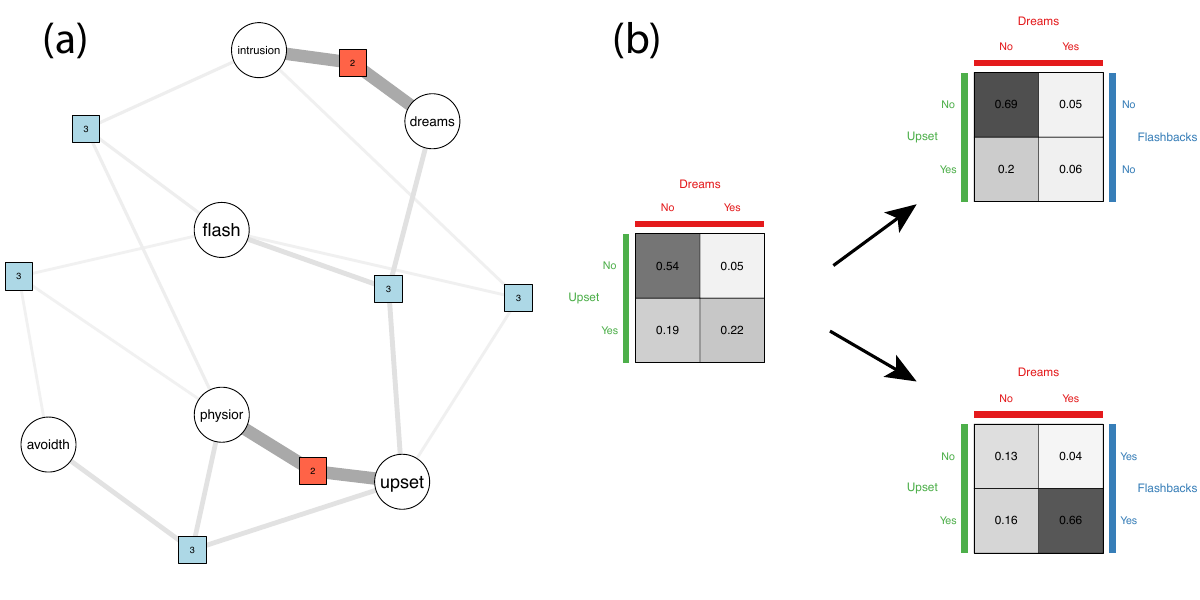}
	
	\caption{(a) Factor graph visualization of the estimated $k=3$ MGM. The circle nodes refer to variables, the quadratic nodes refer to factors over two or three variables. The width of the edges is proportional to the strength of the factor; (b) the marginal sample probability cross-table of \emph{Dreams} and \emph{Upset}, and the same table conditioned on the two states of \emph{Flashbacks}. We see that the relationship between \emph{Dreams} and \emph{Upset} depends on \emph{Flashbacks}}\label{Fig_mgm_applicatio_PTSD}
\end{figure}

\begin{Sinput}
  R> FactorGraph(object = fit_mgmk, 
  +             labels = PTSD_data$names, 
  +             PairwiseAsEdge = FALSE)
\end{Sinput}

which results in Figure \ref{Fig_mgm_applicatio_PTSD} (a). The six circle nodes represent the six variables in the dataset. The red square factor nodes indicate pairwise interactions and the blue square factor nodes indicate 3-way interactions. Each factor node connects to two (pairwise) or three (3-way) variables, indicating an interaction between the respective variables. The width of the edges are proportional to the absolute value of the weight of the corresponding interaction. 

We have a closer look at the 3-way interaction 2-3-4  (\emph{Dreams}, \emph{Flashbacks}, and \emph{Feeling Upset}) in Figure \ref{Fig_mgm_applicatio_PTSD} (b): First look at the marginal probability cross-table of the variables \emph{Dreams} and \emph{Upset}, which shows unequal cell probabilities and hence an interaction between those two variables. Now we condition on the two states of a third variable \emph{Flashbacks} and see that the interaction between \emph{Dreams} and \emph{Upset} considerably depends on whether an individual has \emph{Flashbacks} or not.

Interpreting a $k$-way interaction by interpreting the $k-1$ way interaction for several levels of one the variables $X_j$ in the interaction can be seen as a moderation by $X_j$. In \cite{haslbeck2018moderated} we explain this approach of interpreting $k=3$ interactions as moderation in more detail, and provide further examples for estimating and interpreting higher-order MGMs for the special case of continuous variables.

\subsection{Stationary mixed VAR models}\label{sec_mvar}

In this section, we first show how to estimate a mixed VAR model, compute predictions from it and visualize it based on simulated data. Then we show how to specify and sample from a mixed VAR model in order to generate the data used earlier for estimation. We then fit a mixed VAR model of oder 3 to resting state fMRI data.

\subsubsection{Estimating mixed VAR models}\label{sec_mvar_est}

Here we show how to use the function \code{mvar()} to fit a mixed VAR model to a time series of six variables, consisting of four categorical variables (with 2, 2, 4 and 4 categories) and two Gaussian variables. In the true mVAR model from which the time series was sampled, there are effects of lag order 1 from variable 6 on 5, from 5 on 1 and from 3 on 1. The exact parameterization of these interactions is shown in below in this section, where we create this data set with the function \code{mvarsampler()}.

We provide the data (which is an example dataset automatically loaded with \pkg{mgm}), and specify the type of each variable in \code{type}, where \code{"g"} stands for Gaussian, \code{"p"} for Poisson, and \code{"c"} for categorical. Next, we provide the number of levels for each variable via \code{levels}, where we choose 1 for continuous variables by convention. We specify a lag of order 1 and select the regularization parameters $\lambda$ using the EBIC with tuning parameter $\gamma = .25$:


\begin{Sinput}
  R> fit_mvar <- mvar(data = mvar_data$data, 
  +                  type = c("c", "c", "c", "c", "g", "g"),
  +                  level = c(2, 2, 4, 4, 1, 1),
  +                  lambdaSel = "EBIC",
  +                  lambdaGam = .25, 
  +                  lags = 1)
\end{Sinput}

\code{mvar()} returns a list with several entries: \code{fit_mvar$wadj} is a  $p \times p \times |L|$ array of edge weights, where $|L|$ is the number of specified lags. For example, \code{fit_mvar$wadj[3, 5, 1]} corresponds to the parameter for the crossed lagged effect of 5 on 3 over the first lag specified in \code{lags} (in this example we only specified one lag). \code{fit_mvar$signs} has the same dimension as \code{fit_mvar$wadj} and contains the signs of all parameters, if defined.  \code{fit_mvar$rawlags} contains the full parameterization of the cross-lagged effects. If the mixed VAR model contains only continuous variables, the information in \code{fit_mvar$wadj} and \code{fit_mvar$rawlags} is equivalent. Similarly to \code{mgm()}, the entry  \code{fit_mvar$intercepts} contains all thresholds (intercepts) and \code{fit_mvar$nodemodels} contains the $p$ \pkg{glmnet} models of the  $p$ neighborhood regressions. Here we show the interaction parameters of the fitted VAR model for the single specified lag of order 1:

\begin{Soutput}
  R> round(fit_mvar$wadj[, , 1], 2)
        [,1] [,2] [,3] [,4] [,5] [,6]
  [1,]    0    0 0.33 0.06 0.41 0.00
  [2,]    0    0 0.00 0.00 0.00 0.00
  [3,]    0    0 0.00 0.00 0.00 0.00
  [4,]    0    0 0.00 0.00 0.00 0.00
  [5,]    0    0 0.00 0.00 0.00 0.31
  [6,]    0    0 0.00 0.00 0.00 0.00
\end{Soutput}

The autoregressive effects are on the diagonal and the cross-lagged effects are on the off-diagonal. We use a representation in which columns predict rows, which means that the entry \code{fit_mvar$wadj[5, 3, 1]} corresponds to the cross-lagged effect of 3 on 5 at lag 1. Comparing the estimates with the true cross-lagged effects indicated above, we see that that all three true cross-lagged effects have been recovered and all other effects are correctly set to zero.

The additional arguments that can be provided to \code{mvar()} are similar to the ones in \code{mgm()}: the regularization parameter $\lambda$ can be selected using the EBIC with a specified hyperparameter $\gamma$ or with cross-validation with a specified number of folds. The candidate $\lambda$ sequence is computed as in \code{mgm()} (see Section \ref{sec_mgm}). The $\alpha$ in the elastic-net penalty can be selected with the EBIC or cross validation, similarly to how $\lambda$ is selected. Again similarly to \code{mgm()}, the \code{weights} argument allows to weight observations, \code{binarySign} allows signs for interactions involving binary variables, \code{threshold} defines the type of thresholding (see Section \ref{theory_mgm_estimation}) and \code{overparameterize} allows to choose the preferred type of parameterization of interactions involving categorical variables. For additional input arguments see \code{?mvar}.

In many situations, one fits a VAR model to data that do \emph{not} consist of a sequence of measurements that are equally spaced in time. The reason for this can be (randomly) missing measurements and gaps implied by the measurement process: for instance, in an experience sampling study, individuals may be asked to respond to questions about symptoms 6 times a day at equal time intervals of three hours. A mixed VAR model would then show how the presence of a symptom at a given time point is related to the presence of that and other symptoms at earlier time points (3h ago, 6h ago, etc.). However, because the individual sleeps at night, there are gaps in the time series. If one did not take this information into account, every seventh data point in the time series would represent a lag with the length of the night-gap, whereas the other six are representing a lag of three hours. This problem can be avoided by providing an integer sequence via the argument \code{consec}, which indicates whether measurements are consecutive. For instance if one has a time series with 12 time points (2 days of measurements in the above example), one would provide the vector $c(1, 2, 3, 4, 5, 6, 1, 2, 3, 4, 5, 6)$. If one specifies a lag of order 1, \code{mvar()} then excludes the time step (over night) from measurement 6 to 7. Alternatively one can specify the notification number and the day number via the arguments \code{beepvar} and \code{dayvar}, respectively. Then the \code{consec} variable is computed internally. If a larger number of lags is included, more measurements are excluded accordingly. Information about which cases were excluded as well as the final data matrix used for estimation can be found in \code{mvar$call}. For more details see \code{?mvar} and the application example for the time-varying mVAR model below.

\subsubsection{Making Predictions from mixed VAR model}\label{sec_mvar_pred}

Here we show how to use the \code{predict()} function to compute predictions and nodewise errors from the model estimated in the previous section. We provide the fit object \code{mvar_fit} and the data as arguments:

\begin{Sinput}
  R> pred_mgm <- predict(object = fit_mvar, 
  +                     data = mvar_data$data,
  +                     errorCon = c("RMSE", "R2"),
  +                     errorCat = c("CC", "nCC"))
\end{Sinput}

\code{pred_mgm$call} contains the function call, \code{pred_mgm$predicted} the predicted values for each row in the provided data matrix, and \code{pred_mgm$probabilities} contains the predicted probabilities for categorical variables. \code{pred_mgm$errors} contains a table of nodewise errors.  Similarly to Section \ref{sec_mgm} we specified the Root Mean Squared Error (RMSE) for continuous variables and the (normalized) accuracy for categorical variables:

\begin{Soutput}
  R> pred_mgm$errors
  Variable  RMSE    R2    CC   nCC
  1        1    NA    NA 0.754 0.495
  2        2    NA    NA 0.523 0.000
  3        3    NA    NA 0.302 0.000
  4        4    NA    NA 0.266 0.000
  5        5 0.916 0.157    NA    NA
  6        6 0.998 0.000    NA    NA
\end{Soutput}

Node 1 has the highest normalized accuracy, which makes sense because is predicted by three other nodes at the previous time point. Nodes 2, 3 and 4 have a normalized accuracy of 0, because they are not predicted by any other node. Node 6 has a proportion of explained variance of 0, because it is not predicted by any other node, and node 5 has a nonzero proportion of explained variance because it is predicted by node 6.

One can also provide customary error functions via the \code{errorCon} and \code{errorCat} arguments. For details, see \code{?predict.mgm}.

\subsubsection{Visualizing mixed VAR model}\label{sec_mvar_viz}

We visualize the lagged interaction parameters of the mixed VAR model estimated above together with the nodewise errors computed in the previous section. Specifically, we visualize the proportion of  explained variance for the two continuous variables, and the normalized accuracy for the four categorical variables:

\begin{Sinput}
  R> errors <- c(pred_mgm$errors[1:4, 5], pred_mgm$errors[5:6, 3])
	
  R> qgraph(t(fit_mvar$wadj[, , 1]), 
  +        edge.color = t(fit_mvar$edgecolor[, , 1]), 
  +        pie = errors, 
  +        pieColor = c(rep("tomato", 4), rep("lightblue", 2)),
  +        nodeNames = c(paste0("Categorical; m=",c(2,2,4,4)), 
  +                      rep("Gaussian",2)))
\end{Sinput}

\begin{figure}[H]
	\centering
	\includegraphics[width=.8\textwidth]{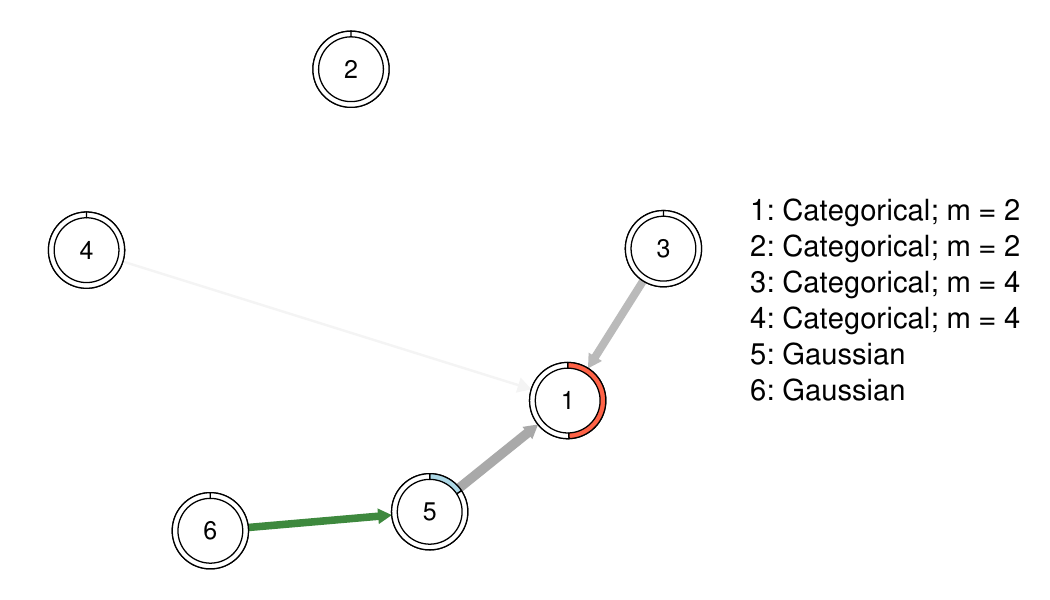}
	\caption{We visualize the lagged interaction parameters of the mixed VAR model estimated above together with the nodewise errors computed in the previous section. Green edges indicate positive relationships. Grey edges indicate that no sign is defined for the pairwise interaction (in the case the interaction involves categorical variables). The width of the edges is proportional to the absolute value of the edge-parameter.}\label{mvar_p6_example}
\end{figure}

We transposed the parameter matrix \code{fit_mvar$wadj[, , 1]} because \code{qgraph()} draws arrows from rows to columns instead of columns to rows, the latter of which is the data structure used in \code{mvar()}. The resulting plot is shown in Figure \ref{mvar_p6_example}.

The green edge indicates a positive linear relationship for the cross-lagged effect from node 6 on node 5. The remaining edges are grey, indicating that no sign is defined. This is because these interactions are defined by several parameters, so no sign can be defined. The width of the edges is proportional to the absolute value of the estimated edge-weights in (the values in \code{fit_mvar$wadj[, , 1]}).

\subsubsection{Sampling from mixed VAR model}\label{sec_mvar_sampling}

We now use the function \code{mvarsampler()} to sample the data set \code{mvar_data} used in the previous section. We specify a model with only one lag of order one and $p = 6$ variables, four categorical (with 2, 2, 4 and 4 categories) and two Gaussians:

\begin{Sinput}
  R> p <- 6
  R> type <- c("c", "c", "c", "c", "g", "g") 
  R> level <- c(2, 2, 4, 4, 1, 1) 
  R> max_level <- max(level)
  R> lags <- 1
  R> n_lags <- length(lags)
\end{Sinput}

Next, we specify the thresholds for each variable. We assign one threshold (intercept) to the Gaussians, and a separate threshold for each of the categories of each of the categorical variables. These thresholds correspond to the first summation in the joint MGM density in Equation~\ref{eq:mixed.joint.full}. In addition, we define a vector indicating the standard deviations of the Gaussian nodes. Note that entries of that vector that do not correspond to Gaussian variables in \code{type} are ignored.

\begin{Sinput}
  R> thresholds <- list()
  R> for(i in 1:p) thresholds[[i]] <- rep(0, level[i])
  R> sds <- rep(1, p)
\end{Sinput}

Finally, we specify the lagged effects in a 5-dimensional $p \times p \times \max\{\text{levels}\}  \times \max\{\text{levels}\} \times |L|$ array, where $|L|$ is the number of lags \code{n_lags}. We first specify the lagged effect from the continuous variable 6 on the continuous variable 5, which consists of a single parameter:

\begin{Sinput}
  R> # Create coefficient array
  R> coefarray <- array(0, dim=c(p, p, max_level, max_level, n_lags))
  R> # Lagged effect: 5 <- 6
  R> coefarray[5, 6, 1, 1, 1] <- .4
\end{Sinput}

We specify two additional lagged effects: one from the categorical variable 3 on the categorical variable 1, which is parameterized by $2 \times 4$ parameters; and one from the continuous variable 5 to the binary variable 1, which is parameterized by $2 \times 1$ parameters.

\begin{Sinput}
  R> # Lagged effect 1 <- 5
  R> coefarray[1, 5, 1:level[1], 1:level[5], 1] <- c(0, 1)
  R> # Lagged effect 1 <- 3
  R> m1 <- matrix(0, nrow=level[2], ncol=level[4])
  R> m1[1,1:2] <- 1
  R> m1[2,3:4] <- 1
  R> coefarray[1, 3, 1:level[2], 1:level[4], 1] <- m1
\end{Sinput}

Finally, all arguments are provided to \code{mvarsampler()}:

\begin{Sinput}
  R> mvar_data <- mvarsampler(coefarray = coefarray,
  +                          lags = lags,
  +                          thresholds = thresholds,
  +                          sds = sds,
  +                          type = type,
  +                          level = level,
  +                          N = 200,
  +                          pbar = TRUE)
\end{Sinput}

These sampled data correspond to the example dataset in \code{mvar_data} we used above to illustrate how to estimate a mVAR model.

\subsubsection{Application: Resting state fMRI data}\label{sec_mvar_application}

We fit an mVAR model with lags 1, 2 and 3 to resting state fMRI data of a single participant. The dataset consists of BOLD measurements of 68 voxels for 240 time points, where the average sampling frequency is 2 seconds \citep[for details see][]{schmittmann2015making}. The data is loaded automatically with the \pkg{mgm} package. All BOLD measurements are modeled as conditional Gaussians, and accordingly we specify the number of levels to be equal to 1 for all variables. We select the regularization parameters $\lambda$ with the EBIC with tuning parameter $\gamma = 0.25$, and we include lags of order 1, 2 and 3.

\begin{Sinput}
  R> rs_mvar <- mvar(data = restingstate_data$data, 
  +                 type = rep("g", 68), 
  +                 level = rep(1, 68), 
  +                 lambdaSel = "EBIC",
  +                 lambdaGam = 0.25,
  +                 lags = c(1, 2, 3))
\end{Sinput}

\begin{figure}[h]
	\centering
	\includegraphics[width=1\textwidth]{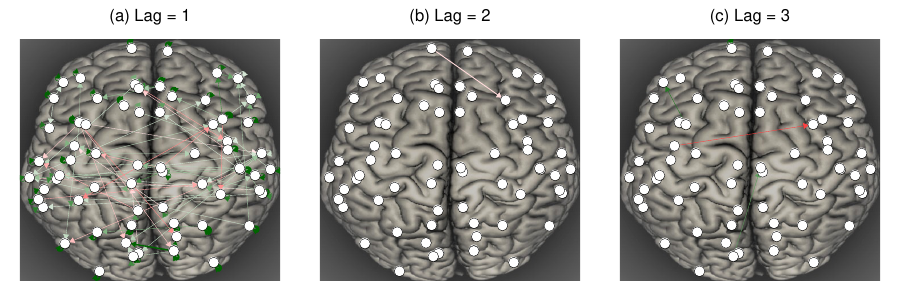}
	\caption{Visualization of the fitted mVAR model, where we depict the parameters separately for each lag. Red edges indicate positive relationships, green edges indicate negative relationships. The width of the edges is proportional to the absolute value of the edge-parameter.}
	\label{mvar_application}
\end{figure}

We visualize the $68 \times 68 \times 3$ interaction parameters of this VAR model in \code{rs_mvar$wadj} in three separate network plots in Figure \ref{mvar_application}, one for each lag. We provide code to reproduce Figure \ref{mvar_application} from the package example data set \code{restingstate_data} in the online supplementary materials and on the Github repository \url{https://github.com/jmbh/mgmDocumentation}.

For the lag of size one, many coefficients are nonzero. In contrast, for the lags of size two and three only few coefficients are nonzero. For a typical fMRI data analysis, this could mean that it is sufficient to fit a VAR model of lag 1 in order to reduce the variance for further analyses.

Similarly to MGMs, the function \code{resample()} can be used to obtain bootstrapped sampling distributions for the parameters of the mVAR model.

\subsection{Time-varying Mixed Graphical Model}\label{sec_tvmgm}

In this section we show how to estimate a time-varying MGM, how to compute predictions from it and how to visualize it.

\subsubsection{Estimating time-varying Mixed Graphical Model}\label{sec_tvmgm_est}

We fit a time-varying MGM to gene expression data used by \cite{gibberd2017regularized}, who took a subset of the data presented by \cite{arbeitman2002gene}. Specifically, we model $p = 150$ gene expressions related to the immune system of D. melanogaster (the fruit fly) measured at $n = 67$ time points across its whole life span. Since $p > n$, this is an example of a high-dimensional estimation problem. Figure \ref{tvmgm_application} (top panel) shows that the 67 measurement are distributed unequally across the time interval.

Estimating the type of time-varying models introduced in Section \ref{theory_timevarying} requires the specification of a bandwidth parameter $\sigma$ that reflects how many time points are combined locally for estimation.  The bandwidth parameter $\sigma$ is the standard deviation of the Gaussian distribution that defines the weighting function (see Section \ref{theory_timevarying}). The empirical time points are normalized to the interval $[0,1]$ and the Gaussian weighting function is defined on this interval. This allows some intuition about which $\sigma$ is appropriate. For example, $\sigma = 2$ implies weights that are close to uniform on the interval $[0,1]$ and therefore gives similar results as the stationary model. This intuition allows to specify a \emph{candidate} $\sigma$-sequence. We select $\sigma$ with the function \code{bwSelect()}, which computes prediction errors on leave-out sets for all candidate $\sigma$-values and selects the $\sigma^*$ that has the lowest mean error. In the next paragraph we describe this approach in detail.

The function \code{bwSelect()} fits time varying models on an equally spaced sequence between the time points $j$ and $n - F + j -1$ of length $J$ with $j \in \{1, 2, \dots, F\}$, where $F$ is the number of folds (times the procedure is repeated) while leaving out (weighting to zero) the time point at which the model is estimated. In a second step, the data at this time point are predicted with the time-varying model and an error measure is computed (RMSE for non-categorical, 0/1-loss for categorical). This procedure is repeated $F$ times. Then we take the arithmetic mean over $J$ estimation points, $p$ variables and $F$ folds. If $J = \frac{n}{F}$, this procedure is equal to a time-stratified $F$-fold cross-validation scheme. We allow to specify $J < \frac{n}{F}$ to save computational cost. $J$ is specified by the argument \code{bwFoldsize} and $F$ is specified by the argument \code{bwFolds}. Selecting the ratio between \code{bwFoldsize} and $n$ corresponds to the problem of selecting the number of folds in cross-validation \citep[see e.g.][]{friedman2001elements}.

For the present illustration we select \code{bwFolds = 5} and \code{bwFoldsize = 5} to keep the computation time short. We provide the candidate $\sigma$-sequence $\{0.1, 0.2, 0.3, 0.4\}$. And we provide all arguments for the time-varying MGM. This is because we repeatedly fit the type of model we want as our final model (then with fixed $\sigma$). We provide the time points of measurements \code{fruitfly_data$timevector} via the argument \code{timepoints} (see Figure \ref{tvmodels_figure_spacing} in Section \ref{theory_timevarying} for an explanation of why one has to provide the time points if they are not equally spaced). Finally, we specify the class of time-varying model \code{modeltype = "mgm"} and pass the arguments \code{k}, \code{threshold} and \code{ruleReg}, to \code{tvmgm()} (see Section \ref{sec_mgm} on \code{mgm()} for a description of these arguments). 

\begin{Sinput}
  R> set.seed(1)
  R> p <- ncol(fruitfly_data$data)
  R> bw_tvmgm <- bwSelect(data = fruitfly_data$data, # Takes ~ 3h
  +                      type = rep("g", p),
  +                      level = rep(1, p),
  +                      bwSeq = c(0.1, 0.2, 0.3, 0.4),
  +                      bwFolds = 5,
  +                      bwFoldsize = 5,
  +                      timepoints = fruitfly_data$timevector,
  +                      modeltype = "mgm", k = 2,
  +                      threshold = "none", ruleReg = "OR")
\end{Sinput}

We would like to know which candidate bandwidth minimized the average prediction error. This information is stored in \code{bw_tvmgm$meanError}:

\begin{Soutput}
  R> round(bw_tvmgm$meanError, 3)
  0.1   0.2   0.3   0.4 
  0.826 0.707 0.630 0.640 
\end{Soutput}

We see that $\sigma = 0.3$ minimizes the error in this dataset. If the smallest/largest candidate $\sigma$ minimized the prediction error, it is advisable to extend the candidate $\sigma$ sequence to smaller/larger values.

After obtaining a reasonable bandwidth for this data set, we can estimate the final time-varying MGM. The estimation points are specified on the unit interval $[0,1]$ to which the provided time scale is normalized internally. We choose 20 equally spaced time points across the time series by setting \code{estpoints = seq(0, 1, length = 20)}. Finally, we specify the above obtained bandwidth with \code{bandwidth = 0.3} and set a random seed to ensure reproducibility.


\begin{Sinput}
  R> set.seed(1)
  R> fit_tvmgm <- tvmgm(data = fruitfly_data$data, 
  +                     type = rep("g", p), 
  +                     level = rep(1, p), 
  +                     timepoints = fruitfly_data$timevector,
  +                     estpoints = seq(0, 1, length = 20),
  +                     k = 2,
  +                     bandwidth = 0.3, 
  +                     threshold = "none", 
  +                     ruleReg = "OR")
\end{Sinput}

The output list in the fit object \code{fit_tvmgm} is similar to the list returned by \code{mgm()}. The difference is that all parameter matrices are now 3 dimensional arrays, with an additional dimension for the estimated time points $t^e \in \mathcal{E}$. For instance, the edge parameters of the pairwise MGM estimated at the third estimation point $t^e=3$ are stored in the matrix \code{fit_tvmgm$pairwise$wadj[, , 3]}. For a a detailed description of all output provided in \code{fit_tvmgm}, see the help file \code{?tvmgm}.

\subsubsection{Making Predictions from time-varying Mixed Graphical Model}\label{sec_tvmgm_pred}

When making predictions with time-varying MGMs, in principle we would need to estimate the time-varying model at the maximum resolution, that is, at every time point. However, this would be computationally expensive: for example, for a time-series of $n = 1000$ time points, we would need to fit 1000 models in order to compute predictions. The \code{predict} method in \pkg{mgm} provides two different options in order to compute predictions and nodewise errors across time, without requiring to estimate $n$ models.

The first option, \code{tvMethod = "weighted"}, computes predictions for each of the $n$ time points from \emph{all} models $t^e \in \mathcal{E}$. It then computes a weighted average over the predictions of all models at each time point. The weight is equal to the weight of the kernel function at $t$ for the respective model estimated at $t^e$. The second option is \code{tvMethod = "closestModel"}, which for each time point determines the closest estimation point $t^e$, and then uses this model for prediction. Accordingly, local nodewise errors are calculated only from the closest model. Note that if one estimates $n$ models at equally spaced time points, this method corresponds to the above described situation of estimating a time-varying model for each time point. 

In order to compute predictions we call the \code{predict()} function and provide the data, the fit object and the desired method to compute predictions. Here we pick \code{tvMethod = "weighted"}:

\begin{Sinput}
  R> pred_tvmgm <- predict(object = fit_tvmgm,
  +                        data = fruitfly_data$data, 
  +                        tvMethod = "weighted")
\end{Sinput}

The output object \code{pred_tvmgm} is a list containing the function call \code{pred_tvmgm$call}, the predicted values \code{pred_tvmgm$predicted} and \code{pred_tvmgm$probabilities} (in the case of categorical variables) computed by the method \code{tvMethod = "weighted"}. \code{pred_tvmgm$true} contains the true data matrix and \code{pred_tvmgm$errors} contains an array of local nodewise error, where the third dimension indicates the estimation points.

\subsubsection{Visualizing time-varying Mixed Graphical Model}\label{sec_tvmgm_viz}

\begin{figure}[H]
	\centering
	\includegraphics[width=1\textwidth]{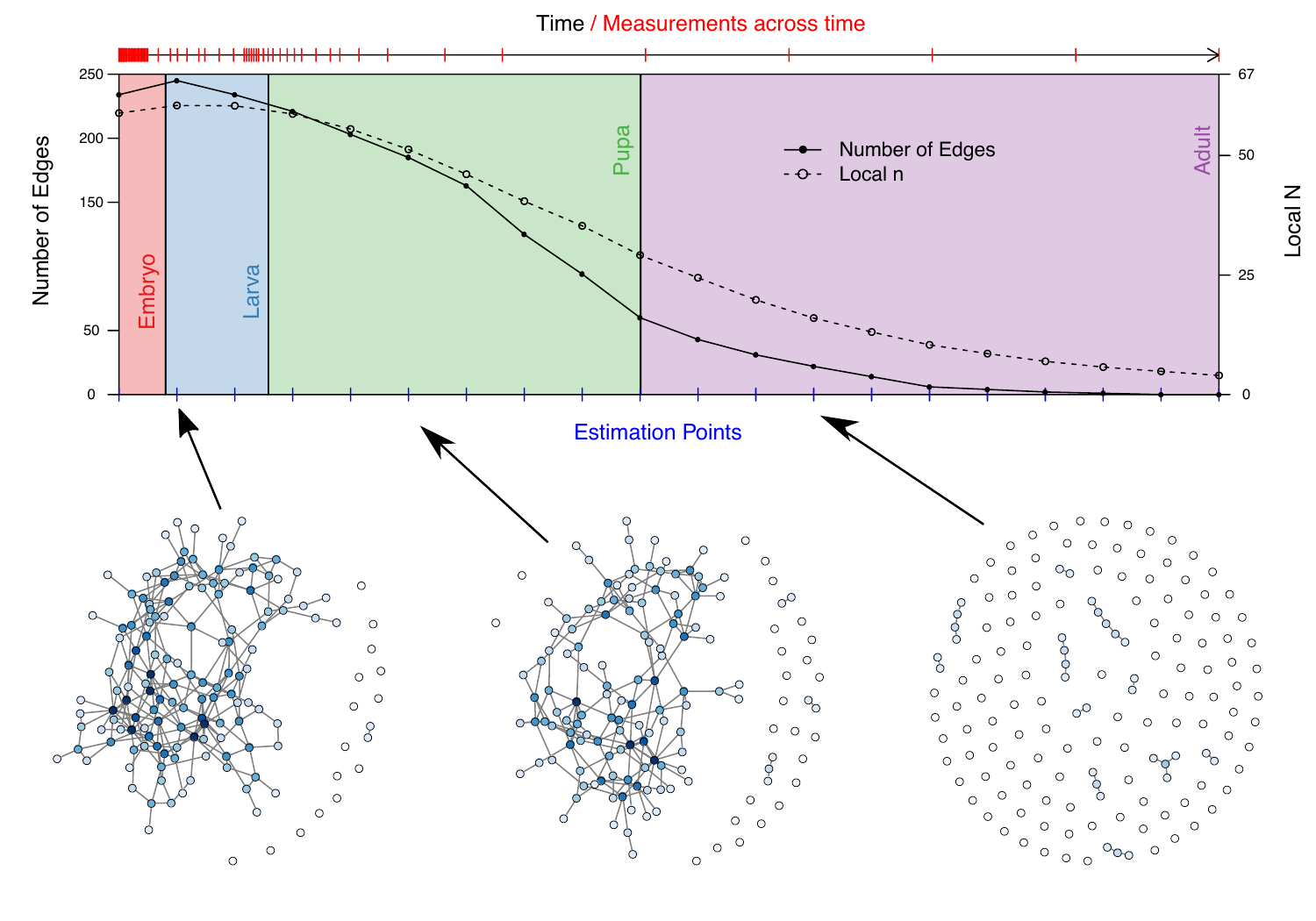}
	\caption{Top panel: the number of estimated edges (solid line) and the local sample size  $n_{\sigma = 0.3, t^e}$ (dashed line) at each estimation point. The red dashes indicate the available measurement on the true time scale. The four colored areas indicate the four stages of the life cycle of the fruit fly. Bottom panel: the undirected network plottet at three different estimation points 2, 6, 13 (with 20 estimation points equally distributed across the 67 time points).}
	\label{tvmgm_application}
\end{figure}

Figure \ref{tvmgm_application} displays several aspects of the time-varying MGM estimated on the fruit-fly data above. The top panel shows the number of edges (solid line) estimated across the time series of 67 measurements, which decreases across the time series. This can be explained by the small number of measurements available at the end of the time series (see red dashes on the time arrow). To make this explicit, we plot $n_{\sigma = 0.3, t^e}$, the used sample size at a given estimation point (see Section \ref{theory_timevarying}). We see that extremely few data points are available in the end of the time series, resulting in a very low sensitivity to detect edges. The lower panel shows the undirected network at the 2nd, 6th and 13th estimation point out of 20 equally spaced estimation points across the whole time series (blue dashes).

While we can interpret the MGM at each estimation in context of the local $n_{\sigma, t^e}$, it is difficult to interpret changes over time, because the sensitivity of the algorithm decreases towards the end of the time series (because less data is available) and hence it is unclear whether edges in the end of the time series are absent in the true model or whether the sensitivity of the algorithm was too low to detect them. This highlights the importance of collecting data with a roughly constant sampling frequency. We provide code to exactly reproduce Figure \ref{tvmgm_application} from the package example data set \code{fruitfly_data} in the online supplementary materials and on the Github repository \url{https://github.com/jmbh/mgmDocumentation}.

\subsubsection{Sampling from time-varying Mixed Graphical Model}\label{sec_tvmgm_sample}

The function \code{tvmgmsampler()} allows to sample from a time-varying MGM. The function input is identical to the input to \code{mgmsampler()}, the sampling function of the stationary MGM described in Section \ref{sec_mgm_sampling}, except that the arguments \code{thresholds}, \code{sds} and \code{interactions} have an additional dimension for time. The number of indices in this additional time dimension defines the length of the time series. Thus, a separate model is specified for each time point in the time series. For details see \code{?tvmgmsampler}.

\subsubsection{Bootstrap Sampling Distributions}\label{sec_tvmgm_resample}

Similarly to stationary MGMs, the function \code{resample()} can be used to obtain bootstrapped sampling distributions for the parameters of the time-varying MGM. The only difference is that we use a block-bootstrapping scheme to ensure that data points remain reasonably distributed across time. The number of blocks can be specified with the argument \code{blocks} in the \code{resample()} function. The larger the number of blocks, the more evenly distributed the bootstrap samples are across the time interval and the higher the similarity between bootstrap samples. Since even distribution across time and low similarity across bootstrap samples is desirable, the number of blocks controls this trade-off. For more details see the help file \code{?resample}.

\subsection{Time-varying mixed VAR model}\label{sec_tvmvar}

We illustrate how to fit a time-varying mixed VAR model on a symptom time series with 51 variables measured on 1478 time points during 238 consecutive days from an individual diagnosed with major depression \citep{wichers2016critical}. The measured variables include questions regarding mood, self-esteem, social interaction, physical activity, events and symptoms of depression (see also legend in Figure \ref{tvmvar_application}). During the measured time interval, a double-blind medication dose reduction was carried out, consisting of a baseline period, the dose reduction, and two post assessment periods (See Figure \ref{tvmvar_application}, the points on the time line correspond to the two dose reductions). For a detailed description of this data set see \cite{kossakowski2017data}.

\subsubsection{Estimating time-varying mixed VAR model}\label{sec_tvmvar_est}

We provide the data, the type (continuous and categorical), and the levels for each variable, all of which are contained in the data list \code{symptom_data} (automatically loaded with \pkg{mgm}), similarly to specifying \code{mvar()}. Next, we provide the day number with \code{dayvar} and the number of notification on each day with \code{beepvar}. Alternatively, one could manually compute a single vector that indicates the consecutiveness of measurements and provide it via the argument \code{consec}. We provide this information because the measurements in this data set are not consecutive, both because of the day-night break in which no measurements are taken and because of randomly missing measurement points. If we did not provide this information, the resulting parameters represent a mixture of effects across different lags and are therefore not interpretable anymore. We explained this in detail in Section \ref{theory_timevarying}. The function \code{tvmvar()} uses this information to fit the model only on rows of the time series for which sufficient previous measurements are available (1 for lag 1, 2 for lag 2, etc.).

In order to fit a time-varying MGM we need to choose an appropriate bandwidth parameter $\sigma$, which determines how many observations close in time we combine in order to estimate a local model (see Section \ref{theory_timevarying}). In Section \ref{sec_tvmgm}, we provided an explanation of how to use \code{bwSelect()} to select an appropriate $\sigma$ using a time-stratified cross validation scheme. Here we choose $\sigma = 0.2$.

We specify a lag of order 1 and via the argument \code{estpoints} we specify that we would like to estimate the model at 20 equally spaced time intervals throughout the time series. We specify the sequence of estimation points on the unit interval $[0,1]$, to which the provided time scale is normalized internally. Finally, we set thresholding \code{threshold = "none"} and set a random seed to ensure reproducibility.

\begin{Sinput}
  > set.seed(1)
  > fit_tvmvar <- tvmvar(data = symptom_data$data, # Takes ~15min
  +                      type = symptom_data$type,
  +                      level = symptom_data$level,
  +                      beepvar = symptom_data$data_time$beepno,
  +                      dayvar = symptom_data$data_time$dayno,
  +                      lags = 1,
  +                      estpoints = seq(0, 1, length = 20),
  +                      bandwidth = 0.2,
  +                      threshold = "none",
  +                      saveData = TRUE)
\end{Sinput}

The output of \code{tvmvar()} is similar to the output of \code{?mvar} described in Section \ref{sec_mvar}. The difference is that all entries have now an additional dimension for estimation points. For example, the entry of the parameter array \code{fit_tvmvar$wadj[4, 9, 2, 15]} indicates the cross lagged effect of 9 on 4 over the second specified lag in \code{lags} at the 15th estimation point. The array \code{fit_tvmvar$signs} has the same dimension and specifies the signs of the parameters in \code{fit_tvmvar$wadj}, if defined. For a discussion of when a sign is defined for an edge-parameter see Section \ref{sec_mgm}. The object \code{fit_tvmvar$intercepts} contains a list with time-varying thresholds/intercepts and \code{fit_tvmvar$tvmodels} contains the models at each of the $|\mathcal{E}|$ estimation points.

We provided the day and notification number of each measurement and \code{tvmvar()} used this information to only include measurements in the model for which sufficient previous measurements are available. By executing the model object in the console, we get the number of measurements that were actually used for estimation:

\begin{Soutput}
  R> fit_tvmvar
  mgm fit-object 
  Model class:  Time-varying mixed Vector Autoregressive (tv-mVAR) model 
  Lags:  1 
  Rows included in VAR design matrix:  876 / 1476 ( 59.35 %) 
  Nodes:  48 
  Estimation points:  20 
\end{Soutput}

We see that for 876 of 1476 measurement points the previous measurement (requirement of lag 1) is available and were therefore used for estimation. If we included lags with higher order the number of usable measurements would become smaller.

\subsubsection{Making Predictions from time-varying mixed VAR model}\label{sec_tvmvar_pred}

In order to compute predictions from the mixed VAR model we have to choose between the two options \code{tvMethod = "weighted"} and \code{tvMethod = "closestModel"}. For a discussion of these two methods see Section \ref{sec_tvmgm} or the help file \code{?predict.mgm}. Next to the fit object we provide the data and information about the consecutiveness of measurements to \code{predict()}:

\begin{Sinput}
  R> # Compute Predictions
  R> pred_tvmvar <- predict(object = fit_tvmvar, 
  +                        data = symptom_data$data,
  +                        tvMethod = "weighted", 
  +                        beepvar = symptom_data$data_time$beepno,
  +                        dayvar = symptom_data$data_time$dayno)
\end{Sinput}

The output object \code{pred_tvmvar} is a list containing the function call \code{pred_tvmgm$call}, the predicted values \code{pred_tvmgm$predicted} and \code{pred_tvmgm$probabilities} (in the case of categorical variables) computed by the method \code{tvMethod = "weighted"}. \code{pred_tvmgm$true} contains the true data matrix, which is useful in the present case, because parts of the (see previous section). Finally, \code{pred_tvmgm$errors} is an array of local nodewise errors, where the third dimension indexes estimation points. For instance,  \code{pred_tvmgm$errors[, , 9]} contains the nodewise errors for estimation point 9.

\subsubsection{Visualizing time-varying mixed VAR model}\label{sec_tvmvar_viz}

\begin{figure}[H]
	\centering
	\includegraphics[width=1\textwidth]{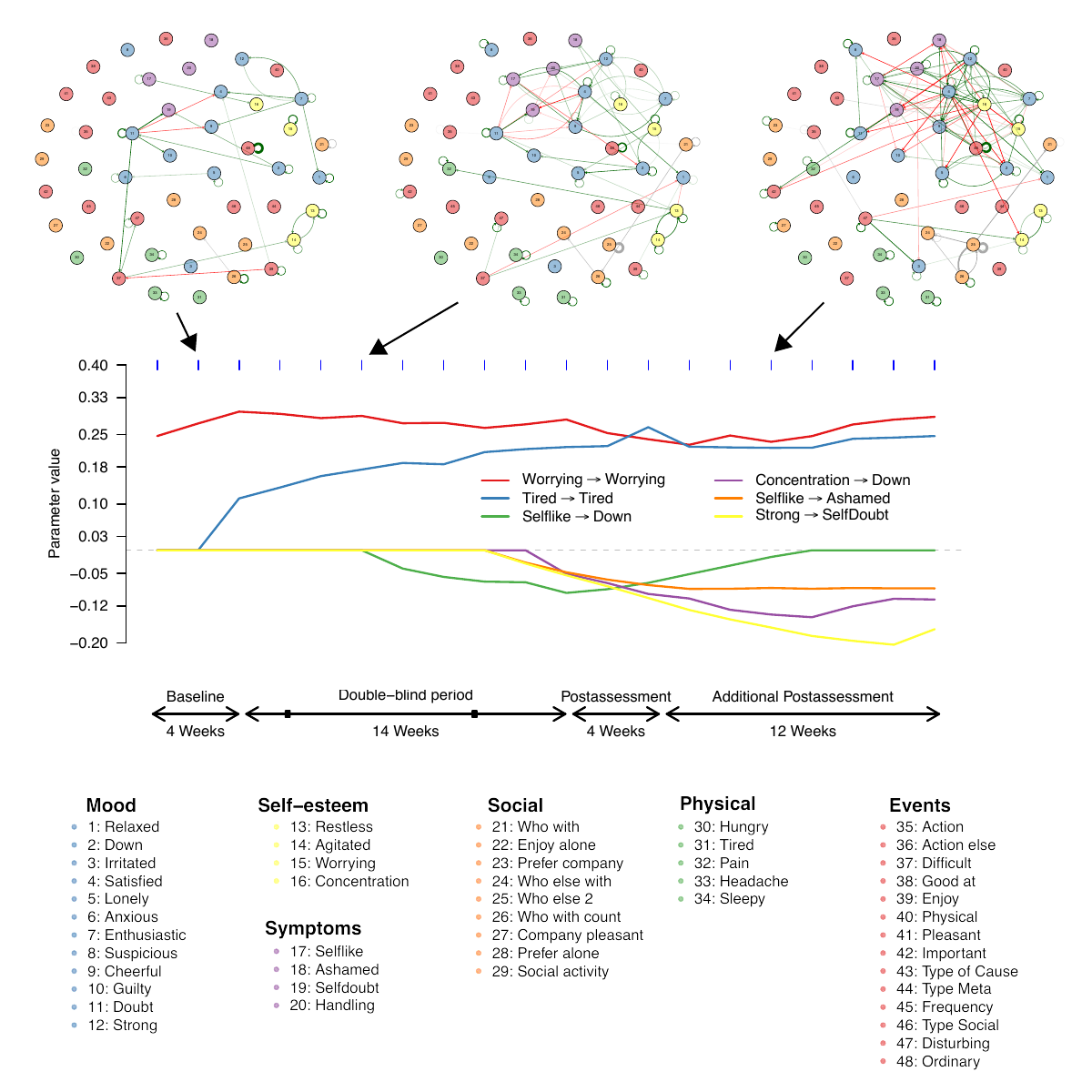}
	\caption{Top row: network visualization of VAR(1) parameters at the estimation points 2, 6, and 16. Green edges indicate positive relationships, red edges indicate negative relationships and grey edges indicate that no sign is defined. The color of the nodes corresponds to the group the variable belongs to (see legend); second row: six autoregressive (e.g., \emph{Worrying}$^{t-1}$ $\rightarrow$ \emph{Worrying}$^{t}$) or cross-lagged effects (e.g., \emph{Selflike}$^{t-1}$ $\rightarrow$ \emph{Down}$^{t}$) depicted as a function of time.} \label{tvmvar_application}
\end{figure}

Figure \ref{tvmvar_application} displays some aspects of the time varying mixed VAR model estimated in the previous section. In the top row of Figure \ref{tvmvar_application} we depict a network plot of the VAR(1) parameters at the estimation points 2, 6, and 16. Green edges indicate positive relationships and red edges indicate negative relationships. Grey edges indicate that no sign is defined, because the edge-weight is a function of several parameters, which is the case for interactions including categorical variables (see Section \ref{sec_mgm}). The width of edges is proportional to the absolute value of the edge-weight. It is evident from the three network plots that the model changes considerably over time which suggests that a stationary model is not appropriate for these data. The second row depicts six autoregressive or cross-lagged effects across the measured time interval. We see that parameters change considerably over time, for instance the autoregressive effect of \emph{Tired} is strong at the beginning of the time series and decreases almost monotonously until the end of the measured time interval.
	
We provide code to exactly reproduce Figure \ref{tvmvar_application} from the example data set \code{symptom_data} in the online supplementary materials and on the Github repository \url{https://github.com/jmbh/mgmDocumentation}.

\subsubsection{Sampling from time-varying mixed VAR model}\label{sec_tvmvar_sample}

The function \code{tvmvarsampler()} allows to sample from a time-varying mVAR model. The function input is identical to the input to \code{mgmsampler()}, the sampling function of the stationary mVAR described in Section \ref{sec_mvar}, except that the arguments \code{thresholds}, \code{sds} and \code{coefarray} have an additional dimension for time. The number of indices in this additional time dimension defines the length of the time series. Thus, a separate model is specified for each time point in the time series. For details see \code{?tvmvarsampler}.

Similarly to time-varying MGMs, the function \code{resample()} allows to obtain bootstrapped sampling distributions for the parameters of time-varying mixed VAR models.

\newpage

\section[CC]{Concluding Comments}

We presented the \pkg{mgm} package which allows to fit stationary and time-varying Mixed Graphical Models and stationary and time-varying mixed Vector Autoregressive Models. In addition to the estimation functions, we provide methods to compute predictions and nodewise errors and assess the stability of estimates via resampling. Furthermore, flexible sampling functions for all model classes allow the user to evaluate the performance of the estimation algorithms in a given situation via simulations. Finally, we provided fully reproducible code examples that illustrate how to use the software package.

The \pkg{mgm} package is under continuous development. We aim to add functions that allow one to inspect higher order interactions in an accessible way. We plan to implement different ways to select tuning parameters ($\lambda$ penalization parameter, $\alpha$ elastic net parameter, $\sigma$ bandwidth parameter), for instance with stability-selection \citep{meinshausen2010stability, liu2010stability}. And we will implement other estimators than $\ell_\alpha$-penalized regression, which might be more appropriate in some situations. Finally, since all estimation algorithms are based on sequential regressions, considerable performance gains can be made by parallelizing the estimation algorithms. 

\bigskip

\centering
\Large \textbf{Acknowledgements}
\normalsize
\flushleft

We would like to thank three anonymous reviewers for their detailed and constructive feedback, and Fabian Dablander and Ois\'in Ryan for their comments on an earlier version of this manuscript.


\bibliography{tbib}{}

\begin{thebibliography}{75}
\newcommand{\enquote}[1]{``#1''}
\providecommand{\natexlab}[1]{#1}
\providecommand{\url}[1]{\texttt{#1}}
\providecommand{\urlprefix}{URL }
\expandafter\ifx\csname urlstyle\endcsname\relax
  \providecommand{\doi}[1]{doi:\discretionary{}{}{}#1}\else
  \providecommand{\doi}{doi:\discretionary{}{}{}\begingroup
  \urlstyle{rm}\Url}\fi
\providecommand{\eprint}[2][]{\url{#2}}

\bibitem[{Agresti(2003)}]{agresti2003categorical}
Agresti A (2003).
\newblock \emph{Categorical data analysis}, volume 482.
\newblock John Wiley \& Sons.

\bibitem[{Albert and Barabasi(2002)}]{Albert_statistical_2002}
Albert R, Barabasi AL (2002).
\newblock \enquote{Statistical Mechanics of Complex Networks.}
\newblock \emph{Reviews of modern physics}, \textbf{74}(1), 47.
\newblock
  \urlprefix\url{http://journals.aps.org/rmp/abstract/10.1103/RevModPhys.74.47}.

\bibitem[{Arbeitman \emph{et~al.}(2002)Arbeitman, Furlong, Imam, Johnson, Null,
  Baker, Krasnow, Scott, Davis, and White}]{arbeitman2002gene}
Arbeitman MN, Furlong EE, Imam F, Johnson E, Null BH, Baker BS, Krasnow MA,
  Scott MP, Davis RW, White KP (2002).
\newblock \enquote{Gene expression during the life cycle of Drosophila
  melanogaster.}
\newblock \emph{Science}, \textbf{297}(5590), 2270--2275.

\bibitem[{Banerjee \emph{et~al.}(2008)Banerjee, El~Ghaoui, and
  d'Aspremont}]{Banerjee_model_2008}
Banerjee O, El~Ghaoui L, d'Aspremont A (2008).
\newblock \enquote{Model Selection through Sparse Maximum Likelihood Estimation
  for Multivariate Gaussian or Binary Data.}
\newblock \emph{The Journal of Machine Learning Research}, \textbf{9},
  485--516.
\newblock \urlprefix\url{http://dl.acm.org/citation.cfm?id=1390696}.

\bibitem[{Begeer \emph{et~al.}(2013)Begeer, Wierda, and
  Venderbosch}]{Begeer_Allemaal_2013}
Begeer S, Wierda M, Venderbosch S (2013).
\newblock \enquote{Allemaal {Autisme}, {Allemaal} {Anders}. {Rapport} {NVA}
  enquête 2013 [{All} {Autism}, {All} {Different}. {Dutch} {Autism} {Society}
  {Survey} 2013].}
\newblock \emph{Bilthoven: NVA.}, p.~83.

\bibitem[{Borsboom and Cramer(2013)}]{Borsboom_Network_2013}
Borsboom D, Cramer AO (2013).
\newblock \enquote{Network {Analysis}: {An} {Integrative} {Approach} to the
  {Structure} of {Psychopathology}.}
\newblock \emph{Annual Review of Clinical Psychology}, \textbf{9}(1), 91--121.
\newblock ISSN 1548-5943, 1548-5951.
\newblock \doi{10.1146/annurev-clinpsy-050212-185608}.
\newblock
  \urlprefix\url{http://www.annualreviews.org/doi/abs/10.1146/annurev-clinpsy-050212-185608}.

\bibitem[{B{\"u}hlmann \emph{et~al.}(2014)B{\"u}hlmann, Kalisch, and
  Meier}]{buhlmann2014high}
B{\"u}hlmann P, Kalisch M, Meier L (2014).
\newblock \enquote{High-dimensional statistics with a view toward applications
  in biology.}
\newblock \emph{Annual Review of Statistics and Its Application}, \textbf{1},
  255--278.

\bibitem[{Casas and Fernandez-Casal(2018)}]{tvReg}
Casas I, Fernandez-Casal R (2018).
\newblock \emph{tvReg: Time-Varying Coefficients Linear Regression for Single
  and Multiple Equations}.
\newblock R package version 0.3.0,
  \urlprefix\url{https://CRAN.R-project.org/package=tvReg}.

\bibitem[{Chen \emph{et~al.}(2015)Chen, Witten
  \emph{et~al.}}]{chen2015selection}
Chen S, Witten DM, \emph{et~al.} (2015).
\newblock \enquote{Selection and estimation for mixed graphical models.}
\newblock \emph{Biometrika}, \textbf{102}(1), 47.

\bibitem[{Chen and He(2015)}]{chen_inference_2015}
Chen X, He Y (2015).
\newblock \enquote{Inference of High-Dimensional Linear Models with
  Time-Varying Coefficients.}
\newblock \emph{arXiv preprint arXiv:1506.03909}.
\newblock \urlprefix\url{http://arxiv.org/abs/1506.03909}.

\bibitem[{Deserno \emph{et~al.}(2016)Deserno, Borsboom, Begeer, and
  Geurts}]{deserno2016multicausal}
Deserno MK, Borsboom D, Begeer S, Geurts HM (2016).
\newblock \enquote{Multicausal systems ask for multicausal approaches: A
  network perspective on subjective well-being in individuals with autism
  spectrum disorder.}
\newblock \emph{Autism}, p. 1362361316660309.

\bibitem[{Efron(1992)}]{efron1992bootstrap}
Efron B (1992).
\newblock \enquote{Bootstrap methods: another look at the jackknife.}
\newblock In \emph{Breakthroughs in statistics}, pp. 569--593. Springer.

\bibitem[{Efron and Tibshirani(1994)}]{efron1994introduction}
Efron B, Tibshirani RJ (1994).
\newblock \emph{An introduction to the bootstrap}.
\newblock CRC press.

\bibitem[{Epskamp(2017)}]{graphicalVAR}
Epskamp S (2017).
\newblock \emph{graphicalVAR: Graphical VAR for Experience Sampling Data}.
\newblock R package version 0.2,
  \urlprefix\url{https://CRAN.R-project.org/package=graphicalVAR}.

\bibitem[{Epskamp \emph{et~al.}(2018)Epskamp, Borsboom, and
  Fried}]{epskamp2018estimating}
Epskamp S, Borsboom D, Fried EI (2018).
\newblock \enquote{Estimating psychological networks and their accuracy: A
  tutorial paper.}
\newblock \emph{Behavior Research Methods}, \textbf{50}(1), 195--212.

\bibitem[{Epskamp \emph{et~al.}(2012)Epskamp, Cramer, Waldorp, Schmittmann, and
  Borsboom}]{qgraph_pkg}
Epskamp S, Cramer AOJ, Waldorp LJ, Schmittmann VD, Borsboom D (2012).
\newblock \enquote{\pkg{qgraph}: Network Visualizations of Relationships in
  Psychometric Data.}
\newblock \emph{Journal of Statistical Software}, \textbf{48}(4), 1--18.
\newblock \urlprefix\url{http://www.jstatsoft.org/v48/i04/}.

\bibitem[{Epskamp \emph{et~al.}(2017)Epskamp, Deserno, and Bringmann}]{mlVAR}
Epskamp S, Deserno MK, Bringmann LF (2017).
\newblock \emph{mlVAR: Multi-Level Vector Autoregression}.
\newblock R package version 0.3.3,
  \urlprefix\url{https://CRAN.R-project.org/package=mlVAR}.

\bibitem[{Foygel and Drton(2010)}]{foygel2010extended}
Foygel R, Drton M (2010).
\newblock \enquote{Extended Bayesian information criteria for Gaussian
  graphical models.}
\newblock In \emph{Advances in neural information processing systems}, pp.
  604--612.

\bibitem[{Foygel and Drton(2014)}]{foygel_high-dimensional_2014}
Foygel R, Drton M (2014).
\newblock \enquote{High-dimensional {Ising} model selection with {Bayesian}
  information criteria.}
\newblock \emph{arXiv preprint arXiv:1403.3374}.
\newblock \urlprefix\url{http://arxiv.org/abs/1403.3374}.

\bibitem[{Friedman \emph{et~al.}(2001)Friedman, Hastie, and
  Tibshirani}]{friedman2001elements}
Friedman J, Hastie T, Tibshirani R (2001).
\newblock \emph{The elements of statistical learning}, volume~1.
\newblock Springer series in statistics Springer, Berlin.

\bibitem[{Friedman \emph{et~al.}(2008)Friedman, Hastie, and
  Tibshirani}]{friedman_sparse_2008}
Friedman J, Hastie T, Tibshirani R (2008).
\newblock \enquote{Sparse Inverse Covariance Estimation with the Graphical
  Lasso.}
\newblock \emph{Biostatistics}, \textbf{9}(3), 432--441.
\newblock ISSN 1465-4644, 1468-4357.
\newblock \doi{10.1093/biostatistics/kxm045}.
\newblock
  \urlprefix\url{http://biostatistics.oxfordjournals.org/cgi/doi/10.1093/biostatistics/kxm045}.

\bibitem[{Friedman \emph{et~al.}(2010)Friedman, Hastie, and
  Tibshirani}]{friedman2010regularization}
Friedman J, Hastie T, Tibshirani R (2010).
\newblock \enquote{Regularization paths for generalized linear models via
  coordinate descent.}
\newblock \emph{Journal of statistical software}, \textbf{33}(1), 1.

\bibitem[{Friedman and Tibshirani(2014)}]{friedman_glasso:_2014}
Friedman J, Tibshirani THaR (2014).
\newblock \enquote{\pkg{glasso}: {Graphical} lasso- estimation of {Gaussian}
  Graphical Models.}
\newblock
  \urlprefix\url{https://cran.r-project.org/web/packages/glasso/index.html}.

\bibitem[{Friedman \emph{et~al.}(2000)Friedman, Linial, Nachman, and
  Pe'er}]{friedmanusing2000}
Friedman N, Linial M, Nachman I, Pe'er D (2000).
\newblock \enquote{Using {Bayesian} {Networks} to {Analyze} {Expression}
  {Data}.}
\newblock \emph{Journal of Computational Biology}, \textbf{7}(3-4), 601--620.
\newblock ISSN 1066-5277.
\newblock \doi{10.1089/106652700750050961}.
\newblock
  \urlprefix\url{http://online.liebertpub.com/doi/abs/10.1089/106652700750050961}.

\bibitem[{Fruchterman and Reingold(1991)}]{fruchterman1991graph}
Fruchterman TM, Reingold EM (1991).
\newblock \enquote{Graph drawing by force-directed placement.}
\newblock \emph{Software: Practice and experience}, \textbf{21}(11),
  1129--1164.

\bibitem[{Ghazalpour \emph{et~al.}(2006)Ghazalpour, Doss, Zhang, Wang,
  Plaisier, Castellanos, Brozell, Schadt, Drake, Lusis, and
  Horvath}]{Ghazalpour_integrating_2006}
Ghazalpour A, Doss S, Zhang B, Wang S, Plaisier C, Castellanos R, Brozell A,
  Schadt EE, Drake TA, Lusis AJ, Horvath S (2006).
\newblock \enquote{Integrating {Genetic} and {Network} {Analysis} to
  {Characterize} {Genes} {Related} to {Mouse} {Weight}.}
\newblock \emph{PLoS Genetics}, \textbf{2}(8), e130.
\newblock ISSN 1553-7390, 1553-7404.
\newblock \doi{10.1371/journal.pgen.0020130}.
\newblock \urlprefix\url{http://dx.plos.org/10.1371/journal.pgen.0020130}.

\bibitem[{Gibberd and Nelson(2015)}]{gibberd2015estimating}
Gibberd AJ, Nelson JD (2015).
\newblock \enquote{Estimating Dynamic Graphical Models from Multivariate
  Time-Series Data.}
\newblock \emph{Proceedings of AALTD 2015}, p.~63.

\bibitem[{Gibberd and Nelson(2017)}]{gibberd2017regularized}
Gibberd AJ, Nelson JD (2017).
\newblock \enquote{Regularized Estimation of Piecewise Constant Gaussian
  Graphical Models: The Group-Fused Graphical Lasso.}
\newblock \emph{Journal of Computational and Graphical Statistics},
  (just-accepted).

\bibitem[{Hamilton(1994)}]{hamiltontime1994}
Hamilton JD (1994).
\newblock \emph{Time {Series} {Analysis}}.
\newblock 1 edition edition. Princeton University Press, Princeton, N.J.
\newblock ISBN 978-0-691-04289-3.

\bibitem[{Haslbeck \emph{et~al.}(2018)Haslbeck, Borsboom, and
  Waldorp}]{haslbeck2018moderated}
Haslbeck J, Borsboom D, Waldorp L (2018).
\newblock \enquote{Moderated Network Models.}
\newblock \emph{archiv.org preprint}.

\bibitem[{Haslbeck and Waldorp(2016)}]{haslbeck2016well}
Haslbeck J, Waldorp LJ (2016).
\newblock \enquote{How well do Network Models predict Future Observations? On
  the Importance of Predictability in Network Models.}
\newblock \emph{arXiv preprint arXiv:1610.09108}.

\bibitem[{Haslbeck and Bringmann(2017)}]{haslbeck2017VAR}
Haslbeck JMB, Bringmann L (2017).
\newblock \enquote{Estimating time-varying VAR models: a comparison of two
  methods.}
\newblock \emph{arXiv preprint: https://arxiv.org/abs/1711.05204}.

\bibitem[{Haslbeck and Waldorp(2017)}]{haslbeck2017mi}
Haslbeck JMB, Waldorp LJ (2017).
\newblock \enquote{Structure Estimation for Mixed Graphical Models in
  High-Dimensional Data.}
\newblock \emph{arXiv preprint: https://arxiv.org/abs/1510.05677}.

\bibitem[{Hastie \emph{et~al.}(2015)Hastie, Tibshirani, and
  Wainwright}]{hastie2015statistical}
Hastie T, Tibshirani R, Wainwright M (2015).
\newblock \emph{Statistical learning with sparsity}.
\newblock CRC press.

\bibitem[{Huang \emph{et~al.}(2010)Huang, Li, Sun, Ye, Fleisher, Wu, Chen, and
  Reiman}]{huang_learning_2010}
Huang S, Li J, Sun L, Ye J, Fleisher A, Wu T, Chen K, Reiman E (2010).
\newblock \enquote{Learning Brain Connectivity of {Alzheimer}'s Disease by
  Sparse Inverse Covariance Estimation.}
\newblock \emph{NeuroImage}, \textbf{50}(3), 935--949.
\newblock ISSN 1053-8119.
\newblock \doi{10.1016/j.neuroimage.2009.12.120}.
\newblock
  \urlprefix\url{http://www.sciencedirect.com/science/article/pii/S1053811909014281}.

\bibitem[{Immer and Gibberd(2017)}]{GraphTime}
Immer A, Gibberd A (2017).
\newblock \enquote{GraphTime.}
\newblock \url{https://github.com/GlooperLabs/GraphTime}.

\bibitem[{Kolar \emph{et~al.}(2010)Kolar, Song, Ahmed, and
  Xing}]{kolar_estimating_2010}
Kolar M, Song L, Ahmed A, Xing EP (2010).
\newblock \enquote{Estimating time-varying networks.}
\newblock \emph{The Annals of Applied Statistics}, \textbf{4}(1), 94--123.
\newblock ISSN 1932-6157.
\newblock \doi{10.1214/09-AOAS308}.
\newblock \urlprefix\url{http://projecteuclid.org/euclid.aoas/1273584449}.

\bibitem[{Kolar and Xing(2009)}]{kolar_sparsistent_2009}
Kolar M, Xing EP (2009).
\newblock \enquote{Sparsistent Estimation of Time-Varying Discrete {Markov}
  Random Fields.}
\newblock \emph{arXiv preprint arXiv:0907.2337}.
\newblock \urlprefix\url{http://arxiv.org/abs/0907.2337}.

\bibitem[{Kolar and Xing(2012)}]{kolar_estimating_2012}
Kolar M, Xing EP (2012).
\newblock \enquote{Estimating Networks with Jumps.}
\newblock \emph{Electronic Journal of Statistics}, \textbf{6}(0), 2069--2106.
\newblock ISSN 1935-7524.
\newblock \doi{10.1214/12-EJS739}.
\newblock \urlprefix\url{http://projecteuclid.org/euclid.ejs/1351865118}.

\bibitem[{Koller and Friedman(2009)}]{koller2009probabilistic}
Koller D, Friedman N (2009).
\newblock \emph{Probabilistic graphical models: principles and techniques}.
\newblock MIT press.

\bibitem[{Kossakowski \emph{et~al.}(2017)Kossakowski, Groot, Haslbeck,
  Borsboom, and Wichers}]{kossakowski2017data}
Kossakowski J, Groot P, Haslbeck J, Borsboom D, Wichers M (2017).
\newblock \enquote{Data from ‘Critical Slowing Down as a Personalized Early
  Warning Signal for Depression’.}
\newblock \emph{Journal of Open Psychology Data}, \textbf{5}(1).

\bibitem[{Lauritzen(1996)}]{Lauritzen_graphical_1996}
Lauritzen SL (1996).
\newblock \emph{Graphical models}.
\newblock Number~17 in Oxford statistical science series. Clarendon Press ;
  Oxford University Press, Oxford, New York.
\newblock ISBN 0-19-852219-3.

\bibitem[{Liu \emph{et~al.}(2010)Liu, Roeder, and Wasserman}]{liu2010stability}
Liu H, Roeder K, Wasserman L (2010).
\newblock \enquote{Stability approach to regularization selection (stars) for
  high dimensional graphical models.}
\newblock In \emph{Advances in neural information processing systems}, pp.
  1432--1440.

\bibitem[{Loh and Wainwright(2012)}]{loh2012structure}
Loh PL, Wainwright MJ (2012).
\newblock \enquote{Structure estimation for discrete graphical models:
  Generalized covariance matrices and their inverses.}
\newblock In \emph{Advances in Neural Information Processing Systems}, pp.
  2087--2095.

\bibitem[{McNally \emph{et~al.}(2015)McNally, Robinaugh, Wu, Wang, Deserno, and
  Borsboom}]{mcnally2015mental}
McNally RJ, Robinaugh DJ, Wu GW, Wang L, Deserno MK, Borsboom D (2015).
\newblock \enquote{Mental disorders as causal systems a network approach to
  posttraumatic stress disorder.}
\newblock \emph{Clinical Psychological Science}, \textbf{3}(6), 836--849.

\bibitem[{Meinshausen and
  B\"{u}hlmann(2006)}]{meinshausen_high-dimensional_2006}
Meinshausen N, B\"{u}hlmann P (2006).
\newblock \enquote{High-dimensional Graphs and Variable Selection with the
  {Lasso}.}
\newblock \emph{The Annals of Statistics}, \textbf{34}(3), 1436--1462.
\newblock ISSN 0090-5364.
\newblock \doi{10.1214/009053606000000281}.
\newblock
  \urlprefix\url{http://projecteuclid.org/Dienst/getRecord?id=euclid.aos/1152540754/}.

\bibitem[{Meinshausen and B{\"u}hlmann(2006)}]{meinshausen2006high}
Meinshausen N, B{\"u}hlmann P (2006).
\newblock \enquote{High-dimensional graphs and variable selection with the
  lasso.}
\newblock \emph{The Annals of Statistics}, pp. 1436--1462.

\bibitem[{Meinshausen and B{\"u}hlmann(2010)}]{meinshausen2010stability}
Meinshausen N, B{\"u}hlmann P (2010).
\newblock \enquote{Stability selection.}
\newblock \emph{Journal of the Royal Statistical Society B}, \textbf{72}(4),
  417--473.

\bibitem[{Monti(2014)}]{monti_2014}
Monti R (2014).
\newblock \enquote{pySINGLE.}
\newblock \url{https://github.com/piomonti/pySINGLE}.

\bibitem[{Monti \emph{et~al.}(2014)Monti, Hellyer, Sharp, Leech,
  Anagnostopoulos, and Montana}]{monti_estimating_2014}
Monti RP, Hellyer P, Sharp D, Leech R, Anagnostopoulos C, Montana G (2014).
\newblock \enquote{Estimating Time-Varying Brain Connectivity Networks from
  Functional {MRI} Time Series.}
\newblock \emph{Neuroimage}, \textbf{103}, 427--443.
\newblock
  \urlprefix\url{http://www.sciencedirect.com/science/article/pii/S1053811914006168}.

\bibitem[{Nelder and Baker(1972)}]{nelder1972generalized}
Nelder JA, Baker RJ (1972).
\newblock \enquote{Generalized linear models.}
\newblock \emph{Encyclopedia of statistical sciences}.

\bibitem[{Nicholson \emph{et~al.}(2017)Nicholson, Matteson, and Bien}]{BigVAR}
Nicholson W, Matteson D, Bien J (2017).
\newblock \emph{BigVAR: Dimension Reduction Methods for Multivariate Time
  Series}.
\newblock R package version 1.0.2,
  \urlprefix\url{https://CRAN.R-project.org/package=BigVAR}.

\bibitem[{Pfaff(2008{\natexlab{a}})}]{pfaff_analysis_2008}
Pfaff B (2008{\natexlab{a}}).
\newblock \emph{Analysis of {Integrated} and {Cointegrated} {Time} {Series}
  with {R}}.
\newblock 2nd edition edition. Springer-Verlag, New York.
\newblock ISBN 978-0-387-75966-1.

\bibitem[{Pfaff(2008{\natexlab{b}})}]{varsPackage}
Pfaff B (2008{\natexlab{b}}).
\newblock \enquote{VAR, SVAR and SVEC Models: Implementation Within {R} Package
  {vars}.}
\newblock \emph{Journal of Statistical Software}, \textbf{27}(4).
\newblock \urlprefix\url{http://www.jstatsoft.org/v27/i04/}.

\bibitem[{Ravikumar \emph{et~al.}(2010)Ravikumar, Wainwright, and
  Lafferty}]{Ravikumar_high-dimensional_2010}
Ravikumar P, Wainwright MJ, Lafferty JD (2010).
\newblock \enquote{High-Dimensional {Ising} Model Selection Using
  L1-Regularized Logistic Regression.}
\newblock \emph{The Annals of Statistics}, \textbf{38}(3), 1287--1319.
\newblock ISSN 0090-5364.
\newblock \doi{10.1214/09-AOS691}.
\newblock \urlprefix\url{http://projecteuclid.org/euclid.aos/1268056617}.

\bibitem[{Schmittmann \emph{et~al.}(2015)Schmittmann, Jahfari, Borsboom, Savi,
  and Waldorp}]{schmittmann2015making}
Schmittmann VD, Jahfari S, Borsboom D, Savi AO, Waldorp LJ (2015).
\newblock \enquote{Making large-scale networks from fMRI data.}
\newblock \emph{PloS one}, \textbf{10}(9), e0129074.

\bibitem[{Schwarz \emph{et~al.}(1978)}]{schwarz1978estimating}
Schwarz G, \emph{et~al.} (1978).
\newblock \enquote{Estimating the dimension of a model.}
\newblock \emph{The Annals of Statistics}, \textbf{6}(2), 461--464.

\bibitem[{Song \emph{et~al.}(2009)Song, Kolar, and Xing}]{song_keller:_2009}
Song L, Kolar M, Xing EP (2009).
\newblock \enquote{{KELLER}: Estimating Time-Varying Interactions between
  Genes.}
\newblock \emph{Bioinformatics}, \textbf{25}(12), i128--i136.
\newblock ISSN 1367-4803, 1460-2059.
\newblock \doi{10.1093/bioinformatics/btp192}.
\newblock
  \urlprefix\url{http://bioinformatics.oxfordjournals.org/cgi/doi/10.1093/bioinformatics/btp192}.

\bibitem[{Tao \emph{et~al.}(2016)Tao, Huang, Wang, Xi, and
  Li}]{tao_multiple_2016}
Tao Q, Huang X, Wang S, Xi X, Li L (2016).
\newblock \enquote{Multiple {Gaussian} {Graphical} {Estimation} with {Jointly}
  {Sparse} {Penalty}.}
\newblock \emph{Signal Processing}.
\newblock ISSN 01651684.
\newblock \doi{10.1016/j.sigpro.2016.03.009}.
\newblock
  \urlprefix\url{http://linkinghub.elsevier.com/retrieve/pii/S0165168416300032}.

\bibitem[{Trip and van Wieringen(2018)}]{trip2018parallel}
Trip DSL, van Wieringen WN (2018).
\newblock \enquote{A parallel algorithm for penalized learning of the
  multivariate exponential family from data of mixed types.}
\newblock \emph{arXiv preprint arXiv:1812.02401}.

\bibitem[{van Borkulo \emph{et~al.}(2014{\natexlab{a}})van Borkulo, Borsboom,
  Epskamp, Blanken, Boschloo, Schoevers, and Waldorp}]{van_borkulo_new_2014}
van Borkulo CD, Borsboom D, Epskamp S, Blanken TF, Boschloo L, Schoevers RA,
  Waldorp LJ (2014{\natexlab{a}}).
\newblock \enquote{A new Method for Constructing Networks from Binary Data.}
\newblock \emph{Scientific Reports}, \textbf{4}.
\newblock ISSN 2045-2322.
\newblock \doi{10.1038/srep05918}.
\newblock \urlprefix\url{http://www.nature.com/doifinder/10.1038/srep05918}.

\bibitem[{van Borkulo \emph{et~al.}(2014{\natexlab{b}})van Borkulo, Epskamp,
  and Robitzsch}]{borkulo_isingfit:_2014}
van Borkulo CD, Epskamp S, Robitzsch wcfA (2014{\natexlab{b}}).
\newblock \enquote{\pkg{IsingFit}: {Fitting} {Ising} Models Using the {eLasso}
  Method.}
\newblock
  \urlprefix\url{https://cran.r-project.org/web/packages/IsingFit/index.html}.

\bibitem[{Wainwright and Jordan(2008)}]{wainwright_graphical_2008}
Wainwright MJ, Jordan MI (2008).
\newblock \enquote{Graphical {Models}, {Exponential} {Families}, and
  {Variational} {Inference}.}
\newblock \emph{Foundations and Trends in Machine Learning}, \textbf{1}(1–2),
  1--305.
\newblock ISSN 1935-8237, 1935-8245.
\newblock \doi{10.1561/2200000001}.
\newblock
  \urlprefix\url{http://www.nowpublishers.com/product.aspx?product=MAL&doi=2200000001}.

\bibitem[{Wainwright \emph{et~al.}(2008)Wainwright, Jordan
  \emph{et~al.}}]{wainwright2008graphical}
Wainwright MJ, Jordan MI, \emph{et~al.} (2008).
\newblock \enquote{Graphical models, exponential families, and variational
  inference.}
\newblock \emph{Foundations and Trends{\textregistered} in Machine Learning},
  \textbf{1}(1--2), 1--305.

\bibitem[{Wan \emph{et~al.}(2015)Wan, Allen, Baker, Yang, Ravikumar, and
  Liu}]{XMRFpackage}
Wan YW, Allen GI, Baker Y, Yang E, Ravikumar P, Liu Z (2015).
\newblock \emph{XMRF: Markov Random Fields for High-Throughput Genetics Data}.
\newblock R package version 1.0,
  \urlprefix\url{https://CRAN.R-project.org/package=XMRF}.

\bibitem[{Wichers \emph{et~al.}(2016)Wichers, Groot, Psychosystems, Group
  \emph{et~al.}}]{wichers2016critical}
Wichers M, Groot PC, Psychosystems E, Group E, \emph{et~al.} (2016).
\newblock \enquote{Critical slowing down as a personalized early warning signal
  for depression.}
\newblock \emph{Psychotherapy and psychosomatics}, \textbf{85}(2), 114--116.

\bibitem[{Wild \emph{et~al.}(2010)Wild, Eichler, Friederich, Hartmann, Zipfel,
  and Herzog}]{wild2010graphical}
Wild B, Eichler M, Friederich HC, Hartmann M, Zipfel S, Herzog W (2010).
\newblock \enquote{A graphical vector autoregressive modelling approach to the
  analysis of electronic diary data.}
\newblock \emph{BMC medical research methodology}, \textbf{10}(1), 28.

\bibitem[{Yang \emph{et~al.}(2014)Yang, Baker, Ravikumar, Allen, and
  Liu}]{yang2014mixed}
Yang E, Baker Y, Ravikumar P, Allen G, Liu Z (2014).
\newblock \enquote{Mixed graphical models via exponential families.}
\newblock In \emph{Artificial Intelligence and Statistics}, pp. 1042--1050.

\bibitem[{Yang \emph{et~al.}(2015)Yang, Ravikumar, Allen, and
  Liu}]{yang2015graphical}
Yang E, Ravikumar P, Allen GI, Liu Z (2015).
\newblock \enquote{Graphical models via univariate exponential family
  distributions.}
\newblock \emph{Journal of Machine Learning Research}, \textbf{16}(1),
  3813--3847.

\bibitem[{Yang \emph{et~al.}(2013)Yang, Ravikumar, Allen, and
  Liu}]{yang2013poisson}
Yang E, Ravikumar PK, Allen GI, Liu Z (2013).
\newblock \enquote{On Poisson graphical models.}
\newblock In \emph{Advances in Neural Information Processing Systems}, pp.
  1718--1726.

\bibitem[{Zhao \emph{et~al.}(2015)Zhao, Li, Liu, Roeder, Lafferty, and
  Wasserman}]{huge_pkg}
Zhao T, Li X, Liu H, Roeder K, Lafferty J, Wasserman L (2015).
\newblock \emph{\pkg{huge}: High-Dimensional Undirected Graph Estimation}.
\newblock \urlprefix\url{http://CRAN.R-project.org/package=huge}.

\bibitem[{Zhao \emph{et~al.}(2012)Zhao, Liu, Roeder, Lafferty, and
  Wasserman}]{zhao_huge_2012}
Zhao T, Liu H, Roeder K, Lafferty J, Wasserman L (2012).
\newblock \enquote{The huge Package for High-Dimensional Undirected Graph
  Estimation in \proglang{R}.}
\newblock \emph{The Journal of Machine Learning Research}, \textbf{13}(1),
  1059--1062.
\newblock \urlprefix\url{http://dl.acm.org/citation.cfm?id=2343681}.

\bibitem[{Zhou \emph{et~al.}(2010{\natexlab{a}})Zhou, Lafferty, and
  Wasserman}]{zhou2010time}
Zhou S, Lafferty J, Wasserman L (2010{\natexlab{a}}).
\newblock \enquote{Time varying undirected graphs.}
\newblock \emph{Machine Learning}, \textbf{80}(2-3), 295--319.

\bibitem[{Zhou \emph{et~al.}(2010{\natexlab{b}})Zhou, Lafferty, and
  Wasserman}]{zhou_time_2010}
Zhou S, Lafferty J, Wasserman L (2010{\natexlab{b}}).
\newblock \enquote{Time Varying Undirected Graphs.}
\newblock \emph{Machine Learning}, \textbf{80}(2-3), 295--319.
\newblock ISSN 0885-6125, 1573-0565.
\newblock \doi{10.1007/s10994-010-5180-0}.
\newblock \urlprefix\url{http://link.springer.com/10.1007/s10994-010-5180-0}.

\bibitem[{Zou and Hastie(2005)}]{zou2005regularization}
Zou H, Hastie T (2005).
\newblock \enquote{Regularization and variable selection via the elastic net.}
\newblock \emph{Journal of the Royal Statistical Society B}, \textbf{67}(2),
  301--320.

\end{thebibliography}

\end{document}